# Computational Insights into Defect Induced Modulation in Electronic Properties of 2D Nitride Monolayers

Shreya G. Sarkar[1], Kuneh Parag Shah[2] and Brahmananda Chakraborty[3,4]*

[1]Accelerator & Pulsed Power Division, Bhabha Atomic Research Centre, Mumbai 400085, India

[2]Department of Physics, Indian Institute of Technology, Roorkee 247667, Uttarakhand, India

[3]Homi Bhabha National Institute, Mumbai 400094, India

[4]High Pressure &Synchrotron Radiation Physics Division, Bhabha Atomic Research Centre, Mumbai 400085, India

*corresponding author

E-mail: brahma@barc.gov.in

## Abstract:

Two-dimensional (2D) nitride materials such as hexagonal boron nitride (h-BN), graphitic carbon nitride (g-$C_3N_4$), and beryllonitrene ($BeN_4$) have emerged as promising candidates for next-generation electronic, optoelectronic, and energy applications due to their unique structural and electronic properties. This study presents a systematic investigation of the effects of vacancy defect, specifically the role of nitrogen and constituent atom vacancies on the electronic properties of these materials. Our findings reveal that the introduction of nitrogen vacancies significantly alters the electronic characteristics of these materials. In h-BN, the presence of a nitrogen monovacancy significantly lowers the work function from 5.97 eV to 3.45 eV, one of the lowest values reported for any 2D material. Additionally, this defect reduces the band gap from 4.6 eV to 0.64 eV, driving the material toward half-metallic behaviour. This is accompanied by the emergence of flat bands near the Fermi level, indicative of strong electron-electron interactions. In g-$C_3N_4$, nitrogen vacancies lead to a decrease in work function and band gap, with double nitrogen vacancies rendering the material nearly metallic. In $BeN_4$, nitrogen vacancies result in minimal charge redistribution and a slight increase in work function, highlighting the material's unique electronic behaviour. These results underscore the potential of vacancy engineering in tuning the electronic properties of 2D nitride materials, offering avenues for the design of materials with tailored work functions and band gaps for applications in optoelectronics, spintronics, and catalysis.

# 1. Introduction

Emerging two-dimensional (2D) materials, including graphene[1], transition metal dichalcogenides[2] (TMDs), black phosphorus[3], and hexagonal boron nitride[4], are revolutionizing electronic[5] and optoelectronic applications[6] due to their exceptional properties. Unlike conventional bulk materials, 2D materials enable the development of ultra-thin, transparent, and flexible devices[7] with improved performance and energy efficiency. The unique band structures of 2D materials gives rise to intriguing phenomena such as the ambipolar field effect[8], the quantum Hall effect at room temperature[9], and ultrahigh electron mobility[10]. Moreover, their compatibility with silicon-based technology and potential for heterostructure engineering[11] make them highly promising for applications in energy generation[12] and storage[13], chemical sensing[14], and spintronics[15].The ongoing synthesis of new 2D materials continues to open up exciting possibilities for miniaturized and multifunctional electronic systems.

Among the emerging two-dimensional (2D) nitride materials, hexagonal boron nitride[16] (h-BN) and other group III nitrides such as 2D GaN, AlN, and InN are attracting significant attention because of their potential applications in electronics. Hexagonal boron nitride (h-BN), with its $sp^2$ planar honeycomb structure similar to graphene, shares many applications with graphene. However, owing to the polar covalent nature of the B-N bond, h-BN exhibits distinct characteristics, such as being a wide band gap semiconductor. This makes it highly suitable as a dielectric and insulating layer[17], as well as an atomically smooth substrate for graphene and other 2D materials, which can enhance carrier mobility and device performance. 2D group III nitride materials also offer high thermal stability[18], wide band gaps, excellent mechanical strength, and superior chemical resistance, making them ideal for high-power and high-frequency electronic devices. As synthesis techniques continue to advance, achieving scalable and defect-free production of 2D nitrides will be essential for their integration into next-generation electronic technologies.

Graphitic carbon nitride (g-$C_3N_4$) is another two-dimensional nitride material from the Group IV family that has been recently synthesized[19]. It is an emerging material with remarkable potential for various applications[20]. As a metal-free semiconductor with a moderate band gap[21] (~2.7 eV), g-$C_3N_4$ exhibits excellent chemical stability and strong visible-light absorption[22] making it a promising candidate for photocatalysis[23], including water splitting and $CO_2$ reduction[24]. Additionally, its layered structure and high nitrogen content provide opportunities for use in energy storage devices[25] where enhanced charge storage and cycling stability are crucial. In electronics, semiconducting nature of g-$C_3N_4$ enables applications in field-effect transistors (FETs) and sensors[26], while its biocompatibility makes it suitable for biomedical[27] and environmental applications.

Beryllium-based 2D materials have recently attracted considerable attention. Two-dimensional beryllium tetranitride (2D $BeN_4$), also known as beryllonitrene is a recently synthesized nitride material. It was first realized experimentally under high-pressure conditions[28], forming a layered structure composed of beryllium atoms coordinated with nitrogen-rich polymeric chains. One of the most striking features of monolayer $BeN_4$ is its Dirac semimetallic character, marked by the presence of anisotropic Dirac cones[29] at the Fermi level. Unlike conventional semiconductors, Dirac materials host massless charge carriers with high mobility and linear energy dispersion, enabling ultra-fast charge transport[30]. This property positions $BeN_4$ as a strong candidate for next-generation nanoelectronic and optoelectronic devices, especially where high-speed, low-dissipation transport is essential. Its unique combination of structural stability, high stiffness[31], and direction-dependent conductivity[32] further enhances its appeal in applications such as field-effect transistors, photodetectors, and spintronic components. First-principles calculations reveal that cutting monolayer $BeN_4$ into zigzag nanoribbons[33] can open a moderate band gap, making them promising for optoelectronic and spintronic applications and high-power devices[34]. Additionally, the strong covalent bonding and high carrier mobility could enhance the performance of field-effect transistors (FETs) and photodetectors. Furthermore, the high nitrogen content of $BeN_4$ and its potential catalytic activity open new avenues for applications in energy storage[35,36] and conversion, including hydrogen evolution reactions (HER) and nitrogen fixation.

Given the remarkable properties of 2D nitrogen-based materials and their numerous potential applications, a crucial area of further research involves understanding how structural modifications, such as vacancy defects, influence their electronic properties. Various defect types, including vacancies, adatoms, and biaxial strain, can significantly impact the electronic behaviour of these materials. Nitrogen doping in 2D materials such as MXenes[37,38] has been shown to effectively modulate their electronic and optoelectronic properties, thereby improving their suitability for optoelectronic applications. Similarly, introducing vacancy defects in nitrogen-based 2D materials could provide a means to fine-tune their electronic properties, potentially optimizing their performance for various applications. In several 2D materials, such as transition metal dichalcogenides (TMDs) and graphene[39], vacancy defects have been shown to alter the band structure, work function[40], carrier concentration, and conductivity, sometimes introducing localized states or enhancing catalytic activity. Vacancy defects occur when atoms are missing from the lattice structure, resulting in notable changes to both electronic and mechanical properties. Studying vacancy defect-induced changes in the electronic properties of 2D nitride materials is crucial for advancing their potential applications in nanoelectronics, optoelectronics, and energy-related technologies.

Previous research have explored various defect engineering approaches such as metal doping[41] and biaxial strain[42] in h-BN, vacancy defects[43] in graphitic carbon nitride[44] for photocatalytic[45] applications, or the influence of vacancy defects on 2D $BeN_4$ monolayer for different applications such as $NH_3$ adsorption and $H_2S$ sensing [46]. However, these studies have largely focused on specific functionalities rather than providing a comprehensive understanding of the overall impact of vacancy defects on electronic properties. A systematic investigation into how different vacancy types influence key electronic properties including work function, band gap, density of states (DOS), and band structure remains largely unexplored. The work function, which determines the minimum energy required to remove an electron from the material's surface, plays a critical role in charge injection efficiency and is essential for applications in field-effect transistors[47] (FETs) and contact engineering. The band gap governs the optical and electronic behaviour of semiconducting 2D materials, making its tunability vital for applications in transistors, photodetectors[48], and optoelectronic devices. The density of states (DOS) provides insights into the availability of electronic states at different energy levels, directly affecting conductivity and carrier transport. Additionally, presence of flat bands characterized by highly localized electronic states with low dispersion can lead to strong electron correlations, exotic quantum phenomena, and even superconductivity[49] in certain 2D materials.

A comprehensive study examining vacancy defects in 2D nitride materials can provide valuable insights into tailoring electronic properties for next-generation devices. In this work, the impact of vacancy defects has been systematically analysed in three distinct two-dimensional (2D) nitride materials: hexagonal boron nitride (h-BN), graphitic carbon nitride ($g\text{-}C_3N_4$) and beryllonitrine ($BeN_4$). Each material represents a different elemental group, offering a diverse platform to explore and compare the role of nitrogen vacancy defects in modifying their electronic properties. The study investigates the effects of nitrogen monovacancies and divacancies in all three materials, focusing on their influence on work function, density of states (DOS), and band structure. Furthermore, vacancy defects involving other constituent elements such as boron in h-BN, barbon in $g\text{-}C_3N_4$ and beryllium in $BeN_4$ are examined to assess their impact on electronic performance. The findings provide critical insights into defect-driven changes, paving the way for the controlled tuning of electronic properties in 2D nitride materials. These insights hold significant potential for applications in nanoelectronics and energy-related technologies.

## 2. **Computational Methods and Models**

The computational analysis was conducted using plane-wave Density Functional Theory (DFT) calculations with the Vienna *Ab initio* Simulation Package (VASP)[50] and the Projector Augmented Wave (PAW) method to solve the Kohn-Sham equations. Energy calculations were performed using the Perdew-Burke-Ernzerhof (PBE) exchange-correlation functional within the generalized gradient approximation (GGA) framework[51]. Spin polarization was incorporated in all calculations, and an energy convergence threshold of $10^{-5}$ eV was set to ensure accuracy. A Monkhorst-Pack k-point grid of 5×5×1 was used for structural optimization, while a denser 9×9×1 grid was applied for density of states and band structure calculations[52]. To prevent periodic interactions perpendicular to the plane, a vacuum of 40 Å was introduced. A plane-wave cutoff energy of 500 eV was used throughout all calculations to ensure convergence. Three distinct species were analysed, with pristine and five different vacancy-defect induced configurations of each species examined for their atomic structure, electronic, and magnetic properties. For the work function calculation, determining the vacuum level was essential which was achieved, by extracting the planar-averaged electrostatic potential along the direction normal to the 2D material surface (z-direction) using the LOCPOT file generated by VASP. A sufficiently large vacuum spacing (25 Å) was used to avoid spurious interactions between periodic images. The Fermi energy was obtained directly from the DOSCAR file. The work function was then computed using the appropriate formula. Bader charge analysis[53] and partial density of states (PDOS) analysis were conducted to investigate charge redistribution and the emergence of new states near the Fermi energy level following the introduction of vacancy defects.

# 3. Simulation Results

## *3.1 Defect Morphologies*

In this study, three different two-dimensional (2D) monolayer materials, namely, hexagonal boron nitride (h-BN), graphitic carbon nitride (g-$C_3N_4$), and beryllonitrene ($BeN_4$) were analysed for various types of vacancy defects. **Figure S1** (Supporting Information) presents a flowchart detailing the various vacancy defects examined in this work. Initially, all pristine structures were simulated, and their properties were validated against existing literature which is shown in the Supporting Information. **Figure S2** displays the pristine crystal structures along with their total density of states (TDOS) plots. For hexagonal boron nitride, a supercell containing 50 atoms with 25 atoms of boron and nitrogen each was used for the simulation, as shown in **Figure S2A**. The optimized pristine structure exhibited a bond length of 1.454 Å, forming six-membered $sp^2$-bonded rings with alternating boron and nitrogen atoms. These bond length values align with previously reported literature[54,55]. The TDOS plot of h-BN, depicted in **Figure S2B**, indicates a wide band gap of 4.60 eV with symmetrical up and down spin states, closely matching previously reported results (4.3–4.75 eV)[56].

For this study, we selected the tri-s-triazine (heptazine)[57] based g-$C_3N_4$, the most stable phase in ambient conditions, which consists of carbon and nitrogen atoms forming polar covalent bonds in a hexagonal lattice structure, as illustrated in **Figure S2C** (Supporting Information). Each Carbon atom bonds with three nitrogen atoms and vice versa. The pristine supercell comprises 56 atoms with 24 carbon and 32 nitrogen atoms. The C-N interatomic distances were found to be 1.331 Å and 1.469 Å, consistent with previously reported values (1.33 Å and 1.46 Å)[58]. **Figure S2D** presents the TDOS plot of 2D triazine(g-$C_3N_4$). **Figure S2E** illustrates the supercell structure of $BeN_4$, which consists of $BeN_4$ pentagons and $Be_2N_4$ hexagons. The optimized lattice parameters were determined to be a = 4.27 Å, b = 3.66 Å, and γ = 64.64°. Each beryllium atom coordinates with four nitrogen atoms, while each nitrogen atom is bonded to three atoms (two nitrogen and one beryllium). The N-N bond length was measured at 1.34 Å (compared to the standard N-N single bond length of 1.45 Å and N=N double bond length of 1.21 Å), while the Be-N bond length was 1.45 Å, aligning well with existing literature[59]. The pristine $BeN_4$ supercell contains 45 atoms comprising of 9 beryllium and 36 nitrogen atoms. **Figure S2F** presents the TDOS plots for the defect structures of $BeN_4$. The TDOS plot of the pristine structure indicates zero band gap, consistent with the theoretical predictions of Bykov et al. which suggest that the $BeN_4$ monolayer exhibits semimetallic properties with Dirac points in its electronic band structure at the Fermi level. This finding is further supported by our simulation results.

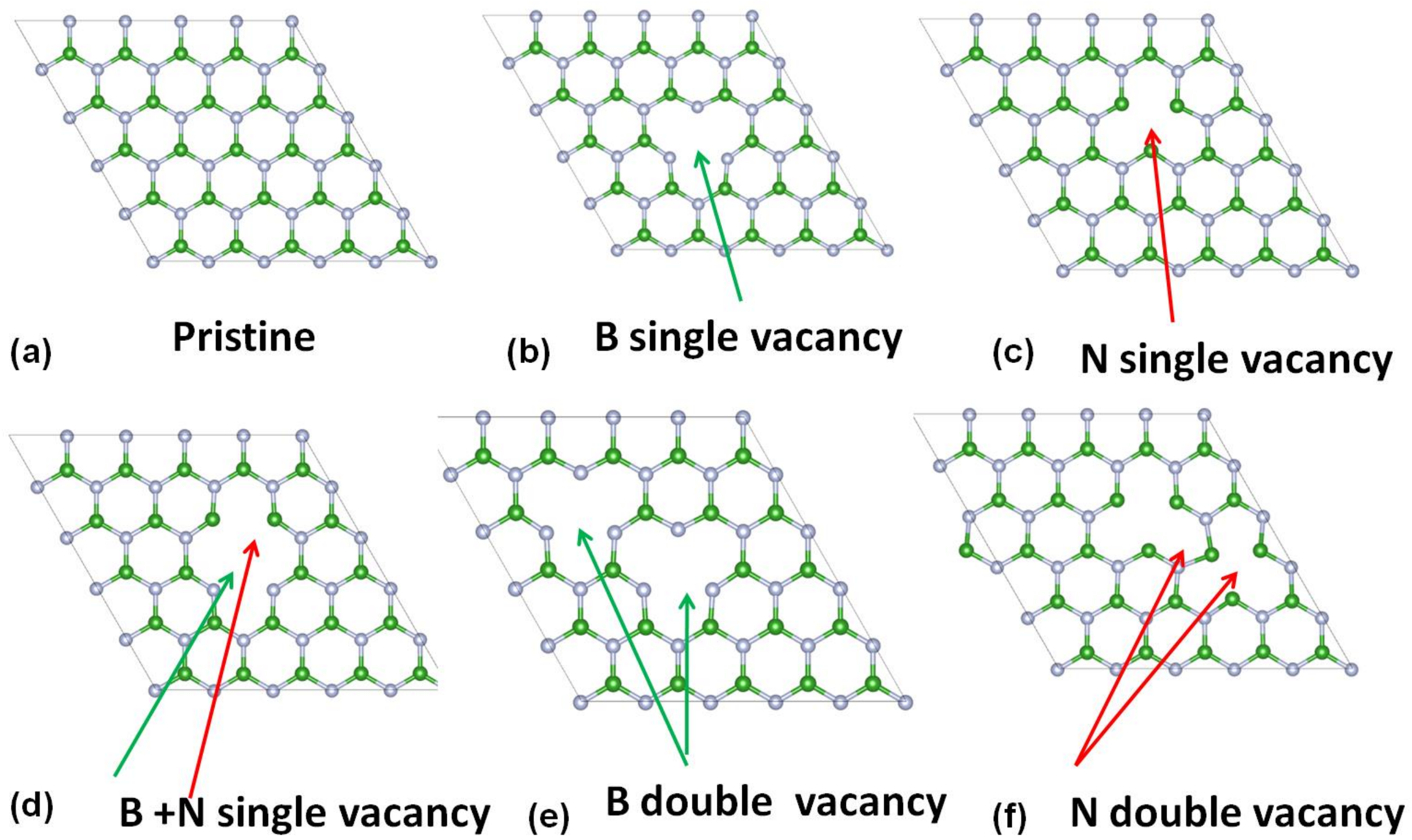

**Figure 1.** Optimized supercell structures of hexagonal boron nitride (h-BN). boron atoms are shown in green and nitrogen atoms in transparent color.(a) **BNP**: pristine h-BN;(b) **BND1**: h-BN with a single boron vacancy; (c) **BND2**: h-BN with a single nitrogen vacancy;(d) **BND3**: h-BN with a boron–nitrogen divacancy; (e) **BND4**: h-BN with a double boron vacancy; (f) **BND5**: h-BN with a nitrogen divacancy.

After verifying all the pristine structures, five types of vacancy defects were introduced into each structure, as illustrated in **Figures 1** and **Figures S3** and **S4**. The defect formation process was carried out in two steps. First, vacancies were created by selectively removing atoms from the relaxed pristine structure according to the specific vacancy type, ensuring the defect density remained within a controlled range. Next, the defective structures were relaxed until their energies converged. In the case of monovacancy defects, a single atom was removed, whereas for double vacancy defects, either two atoms of the same element or a combination of one atom from each constituent element were removed. The resulting relaxed structures were then used for further analysis. For boron nitride, a single vacancy corresponds to a defect density of 2%, while a double vacancy results in 4%. Similarly, the defect densities for graphitic carbon nitride are 1.78% for a single vacancy and 3.57% for a double vacancy, while for beryllonitrene, they are 2.22% and 4.44%, respectively. All these defect densities fall within experimentally feasible ranges.

**Figure 1** illustrates the atomic structures of vacancy defects in h-BN after relaxation. **Figure 1a** represents the pristine structure, while **Figures 1b–1c** show monovacancy defects for boron (B) and nitrogen (N) respectively. **Figure 1d** represents a combination of both B and N vacancy whereas **Figure 1e** and **1f** depict double vacancy defects for boron and nitrogen atoms respectively. In the figures, boron atoms are shown in green, and nitrogen atoms in grey. Red arrows indicate nitrogen vacancy defect sites, while green arrows highlight vacancy defects for boron/carbon/beryllium in **Figure 1** ,**S3** and **S4** (Supporting Information). The introduction of vacancy defects causes surrounding atoms to shift and adjust to stabilize the structure, yet the overall periodicity remains intact, as evident from the figures. **Table S1** presents the changes in bond lengths after structural relaxation. Simulation results indicate that the bond length increases for nitrogen vacancy defects but decreases for boron vacancy defects compared to the pristine structure.

**Figures S3** and **S4** illustrate the atomic structures of vacancy defects in g-$C_3N_4$ and $BeN_4$, respectively, after relaxation. In the figures, carbon atoms are represented in brown, beryllium (Be) atoms in light green, and nitrogen atoms in grey. **Table S1** (Supporting Information) presents the changes in bond lengths following relaxation. For g-$C_3N_4$, two distinct bond C-N and C=N bonds are observed, one associated with the hexagonal ring and the other corresponding to a free C=N bond. In the case of a nitrogen vacancy defect, the C=N bond within the hexagonal ring shortens, whereas for a carbon vacancy, this bond length increases. In the $BeN_4$ structure, both Be-N ionic-covalent and N-N covalent bond lengths decrease when a nitrogen vacancy is introduced. However, for a beryllium vacancy, the Be-N bond length remains unchanged compared to the pristine structure.

### *3.2 Formation energy of the defective structure*

The formation energy of a modified structure is calculated using the following equation:

$$E_{formation} = E_{\text{mod}ified} - E_{pristine} + \sum \mu_i n_i \qquad (1)$$

In Equation 1, $E_{formation}$ stands for formation energy typically in eV, $E_{modified}$ stands for the total energy of the modified structure, $E_{pristine}$ stands for Energy of the pristine structure, $\mu_i$ stands for the chemical potential of the atom removed or added and $n_i$ stands for number of atoms of that has been removed or added with appropriate sign. Previous studies have calculated the chemical potential of carbon using a single carbon atom in a carbon compound[60]. We have utilised a 32 atom graphene sheet under the same computational settings and found out the total energy after structural relaxation to be -295.6eV. Therefore, the chemical potential of C is calculated as $\mu_C$ = -9.237eV, which is close to -9.267eV as reported in literature[61]. Further, using the total energy of the structurally optimised pristine g-$C_3N_4$ (-472.3098eV), the chemical potential of N atom was back-calculated ($\mu_N$ = -7.831eV). In similar fashion, using the total energy values of pristine h-BN (-441.7907 eV) and $BeN_4$ (-329.2927 eV), chemical potential for boron ($\mu_B$ = -9.841 eV) and beryllium ($\mu_{Be}$ = -5.264 eV) were calculated.

**Table1**: Formation energy of structure under consideration for simulation

| **Material** | **Defect Type** | **Formation energy ($E_{formation}$) in eV** |
|---|---|---|
| h-BN | Pristine | 0 |
| | N mono vacancy | 7.86 |
| | B mono vacancy | 7.5 |
| | B+N | 9.62 |
| | N Double vacancy | 15.42 |
| | B Double Vacancy | 14.81 |
| g- $C_3N_4$ | Pristine | 0 |
| | N mono vacancy | 2.49 |
| | C mono vacancy | 2.17 |
| | C+N | 4.69 |
| | N Double vacancy | 5.05 |
| | C Double Vacancy | 2.24 |
| $BeN_4$ | Pristine | 0 |
| | N mono vacancy | 2.47 |
| | Be mono vacancy | 4.39 |
| | Be+N | 5.72 |
| | N Double vacancy | 4.86 |
| | Be Double Vacancy | 9.11 |

The formation energy determines the most energetically favourable vacancy structures among those considered for simulation. The formation energies of the individual structures under investigation are presented in **Table 1** and **Figure S5**. Among the three different 2D nitrides studied, the defective h-BN structure exhibits the highest formation energy, while graphitic carbon nitride has the lowest formation energy, making the defective structures more likely to form. In all cases, monovacancy defects have lower formation energy compared to their double vacancy counterparts, making them more energetically favourable. For both h-BN and g-$C_3N_4$, the formation energies of atomic vacancy defects are similar. However, in $BeN_4$, nitrogen vacancy defects are more favourable than beryllium vacancy defects.

### *3.3 Work Function of the defective structure*

In this work, we have simulated the change in work function with vacancy defect. The Work Function is the minimum amount of energy required to remove an electron from the surface of a material to a point in vacuum just outside the material. The Work function of a material (Φ) can be calculated using the following equation:

$$\Phi = V - E_F \quad (2)$$

Where $V$ stands for the vacuum electrostatic potential and $E_F$ stands for the Fermi energy level.

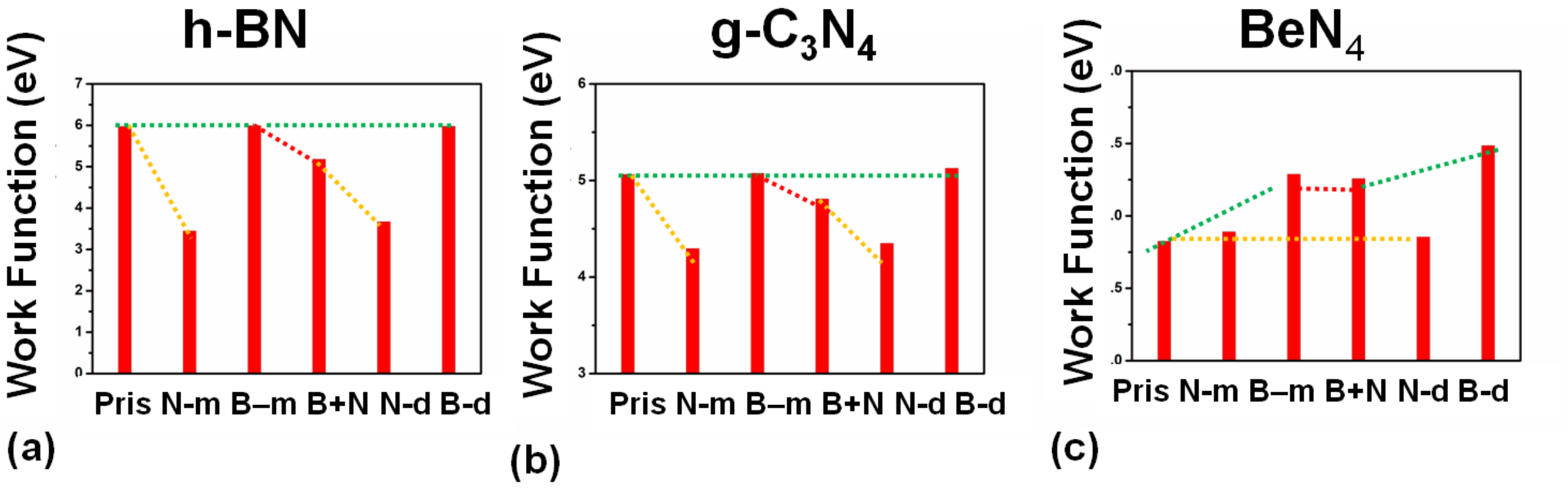

**Figure 2**: Modulation of work function values. From left to right: $M_xN_\gamma$ pristine; $M_xN_\gamma$ with an N vacancy; $M_xN_\gamma$ with an M atom vacancy; $M_xN_\gamma$ with an N and M atom double vacancy; $M_xN_\gamma$ with an N double vacancy; $M_xN_\gamma$ with an M double vacancy, where M stands for (a) boron in h-BN, (b) carbon in g-$C_3N_4$, and (c) beryllium in $BeN_4$, respectively. The green dotted line indicates the change in work function due to an M-type vacancy, the yellow line indicates the change in work function due to an N vacancy, and the red line indicates the change in work function due to both types of vacancies.

**Figure 2** and **Figures S6-S8** (Supporting Information) illustrate the variation in the work function of the examined structures due to different types and densities of atomic vacancy defects. The work function values for each structure, along with their percentage change from the pristine structures, are presented in **Table 2**. **Figure 2a** specifically shows the work function changes in the h-BN structure. It is evident from the figure that the introduction of a nitrogen vacancy leads to a drastic decrease in the work function, whereas a boron vacancy has no significant effect on work function. When both boron and nitrogen vacancies are present, a reduction in work function is observed, though the percentage change is smaller compared to a nitrogen vacancy defect alone. Notably, a nitrogen monovacancy in hexagonal boron nitride results in a significant 1.53 eV (-42%) reduction in the work function. The work function value of h-BN with a nitrogen mono vacancy defect is 3.4 eV which is among the lowest reported for a 2D structure (TableS3, Supporting Information). As the defect density increases, the work function remains largely unchanged from its mono vacancy defect values and even shows a slight increase compared to a single nitrogen atomic vacancy, as observed in **Figure 2a**.

**Table 2**: Work Function of the Pristine and vacancy defect structures of h-BN, g-$C_3N_4$ and $BeN_4$

| Material | Defect Type | Vacuum potential (eV) | Fermi Level (eV) | Work Function (eV) | % change in work function |
|---|---|---|---|---|---|
| h-BN | Pristine | 0.764 | -5.21 | 5.97 | 0 |
| | N mono vacancy | 0.763 | -2.68 | 3.45 | -42.21 |
| | B mono vacancy | 0.75 | -5.24 | 5.99 | 0.33 |
| | B+N | 0.753 | -4.43 | 5.18 | -13.23 |
| | N Double vacancy | 0.75 | -2.92 | 3.67 | -38.53 |
| | B Double Vacancy | 0.745 | -5.23 | 5.98 | 0.17 |
| g- $C_3N_4$ | Pristine | 0.749 | -4.31 | 5.06 | 0 |
| | N mono vacancy | 0.742 | -3.55 | 4.3 | -15.02 |
| | C mono vacancy | 0.742 | -4.33 | 5.08 | 0.39 |
| | C+N | 0.736 | -4.07 | 4.81 | -4.94 |
| | N Double vacancy | 0.735 | -3.61 | 4.35 | -14.03 |
| | C Double Vacancy | 0.734 | -4.395 | 5.13 | 1.38 |
| $BeN_4$ | Pristine | 1.45 | -2.374 | 3.83 | 0 |
| | N mono vacancy | 1.41 | -2.48 | 3.89 | 1.56 |
| | Be mono vacancy | 1.39 | -2.89 | 4.29 | 12.01 |
| | Be+N | 1.37 | -2.88 | 4.26 | 11.22 |
| | N Double vacancy | 1.36 | -2.49 | 3.86 | 0.78 |
| | Be Double Vacancy | 1.33 | -3.15 | 4.49 | 17.23 |

Similarly, in graphitic carbon nitride (g-$C_3N_4$), the change in work function follows a trend similar to that of h-BN. **Figure 2b** illustrates the variation in work function for g-$C_3N_4$, where a nitrogen vacancy leads to a decrease, while a carbon vacancy has no significant effect. However, the percentage change in work function for g-$C_3N_4$ is less pronounced compared to h-BN. The vacuum potential level in the defective structure remains largely unchanged despite the creation of vacancies. However, in the case of a nitrogen vacancy defect, the electrons from the unsaturated dangling bonds introduces the localized electronic states on neighbouring atoms within the band gap. These localised states may shift the Fermi level upward (**Table 2**) to accommodate unpaired valence electrons thereby reducing the overall work function. In contrast, beryllonitrene deviates from this trend, as shown in **Figure 2c**. In this material, a nitrogen vacancy results in either a negligible increase or an almost unchanged work function compared to the pristine structure, whereas a beryllium vacancy leads to an increase in work function. This behaviour can be attributed to the higher number of nitrogen atoms in $BeN_4$ compared to the other two structures. The impact of nitrogen vacancies is less significant in $BeN_4$ samples. In conclusion, the simulation results suggest that creation of nitrogen vacancies acts as an effective mechanism for reducing the work function in 2D nitride materials. Low work function materials enable a range of advanced electronic and optoelectronic applications. Work function reduction is critically important for devices such as thermionic emitters[62], electron guns, and vacuum microelectronics, where high electron output at lower operating voltages or temperatures is desired. In field emission devices[63], a reduced work function lowers the threshold voltage required for electron emission, enhancing efficiency and enabling cold cathode technologies such as flat-panel displays, X-ray sources, and electron microscopes. In photocathodes and photoelectron devices, reduced work function materials improve quantum efficiency[64] and enable operation under lower-energy light sources. Additionally, in Schottky contact engineering, tuning the work function allows for optimized barrier heights at metal-semiconductor interfaces, facilitating improved carrier injection or extraction, which is critical for high-speed diodes, transistors, and photodetectors.

### *3.4 **The density of state (TDOS) and band gap of the defective structure***

**Figures 3** to **5** illustrate the total density of states (TDOS) as a function of energy for pristine and defective structures of h-BN, g-$C_3N_4$, and $BeN_4$, respectively. The x-axis is shifted by the Fermi energy of each structure. **Figure 6** presents the simulated band gap values of these structures and highlights the changes caused by different types of vacancy defects. **Figure 3a** shows the TDOS plot for pristine h-BN, which has a band gap of 4.6 eV. This wide band gap makes h-BN highly effective as an insulator in heterostructures. However, our simulation results (**Figures 3b–3f**) reveal a significant reduction in the band gap when vacancy defects are introduced. This reduction occurs due to the formation of new states near the Fermi level within the band gap region of the vacancy defect structure which narrows the overall gap as can be seen from the TDOS plots. Notably, the decrease in band gap is more pronounced in structures with boron (B) vacancies compared to those with nitrogen (N) vacancies (**Figures 3c**, **3f**, and **6a**). The TDOS plot depicted in **Figure 3f** for h-BN with a double B vacancy, demonstrates half-metallicity in the defective structure, as only spin-down channel states appear to be present near the Fermi level. Additionally, the TDOS plots indicate spin polarization asymmetry, leading to a magnetic moment in all monovacancy-defective structures and in the structure along with a double B vacancy (**Figure S9**, **Table S2**). In contrast, for h-BN with a double N vacancy defect, the symmetry is restored, resulting in a net magnetic moment of zero.

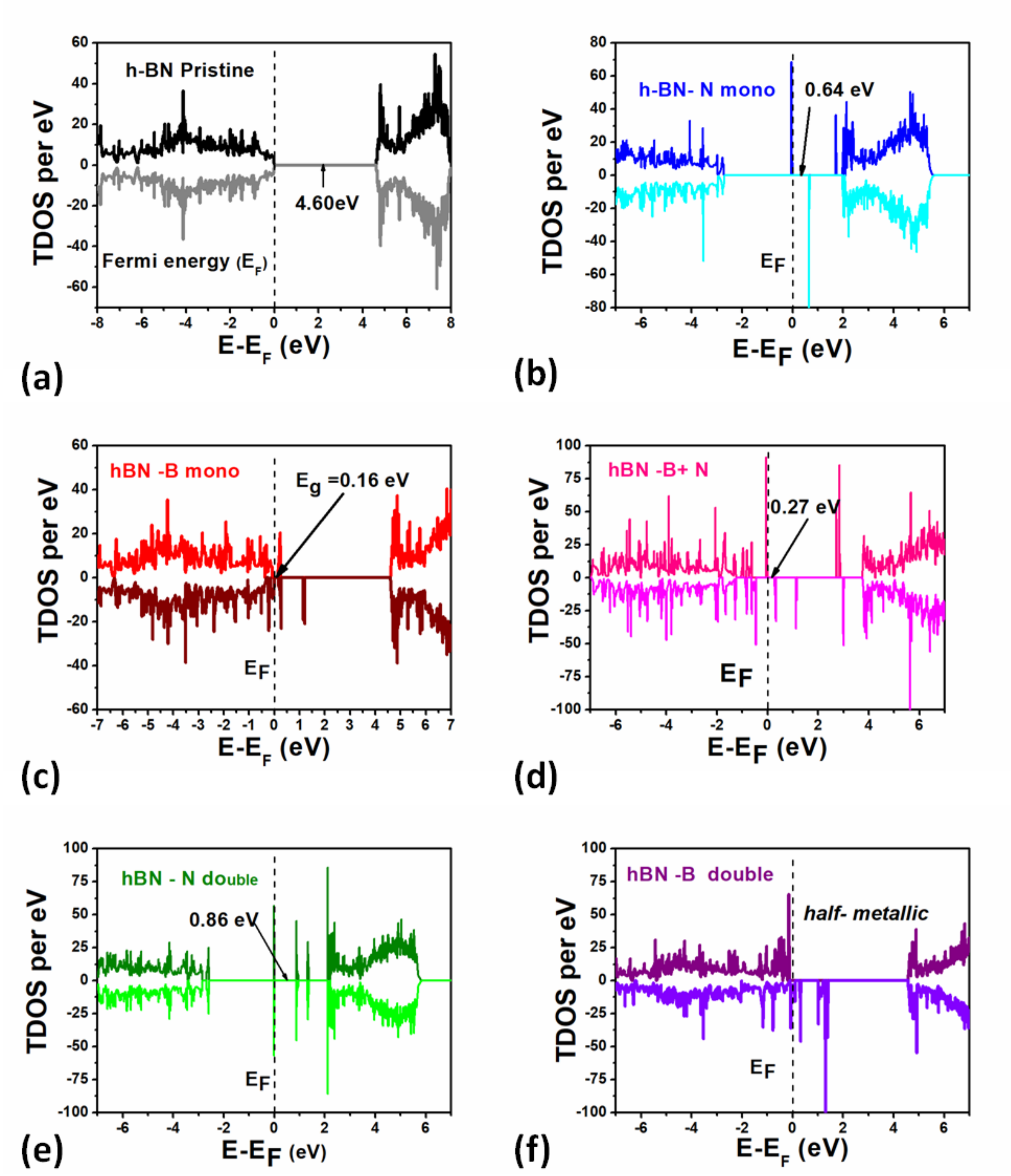

**Figure 3.** Total density of states (TDOS) plots for pristine and defective h-BN structures. (a) Pristine h-BN (solid black: spin-up, solid grey: spin-down),(b) h-BN with nitrogen monovacancy (solid blue: spin-up, solid cyan: spin-down),(c) h-BN with boron monovacancy (solid red: spin-up, solid brown: spin-down),(d) h-BN with boron + nitrogen vacancy (solid pink: spin-up, solid magenta: spin-down),(e) h-BN with nitrogen double vacancy (solid dark green: spin-up, solid light green: spin-down),(f) h-BN with boron double vacancy (solid purple: spin-up, solid violet: spin-down).

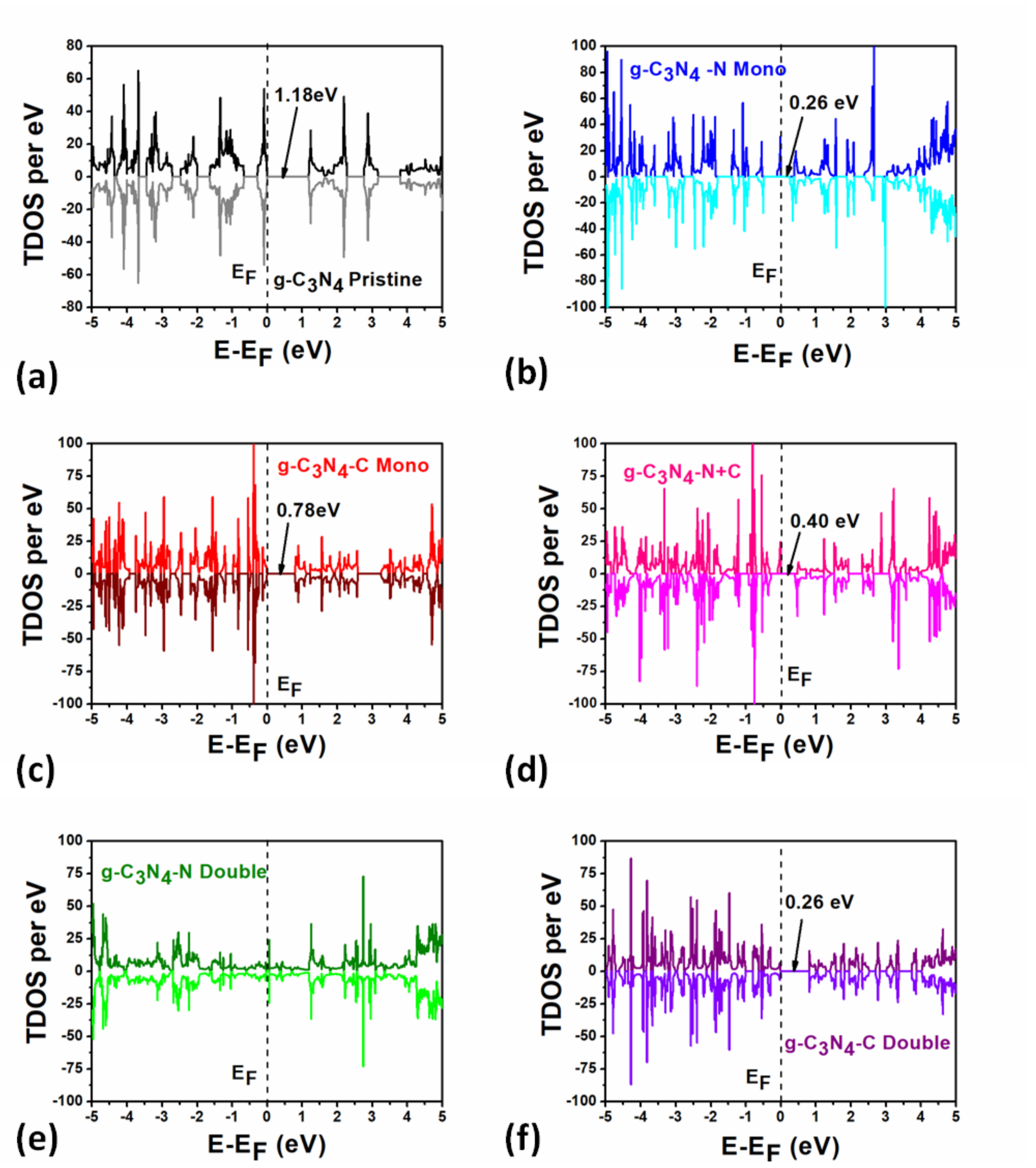

**Figure 4**. Total density of states (TDOS) plots for pristine and defective graphitic carbon nitride ($g-C_3N_4$) structures. (a) Pristine $g-C_3N_4$ (solid black: spin-up, solid grey: spin-down),(b) $g-C_3N_4$ with nitrogen monovacancy (solid blue: spin-up, solid cyan: spin-down),(c) $g-C_3N_4$ with carbon monovacancy (solid red: spin-up, solid brown: spin-down),(d) $g-C_3N_4$ with carbon + nitrogen vacancy (solid pink: spin-up, solid magenta: spin-down),(e) $g-C_3N_4$ with nitrogen double vacancy (solid dark green: spin-up, solid light green: spin-down),(f) $g-C_3N_4$ with carbon double vacancy (solid purple: spin-up, solid violet: spin-down).

**Figures 4a**–**4f** illustrate the TDOS plots for both pristine and defective structures of graphitic carbon nitride (g-$C_3N_4$). As shown in **Figure 4a,** the pristine structure has a band gap of 1.18 eV, which is comparable to that of silicon. Since visible light spans an energy range of 1.67 eV to 3.22 eV, g-$C_3N_4$ can absorb the entire visible spectrum, making it a promising material for photocatalytic applications. Similar to boron nitride, the introduction of vacancies generates additional states within the band gap, leading to a significant reduction in its size (**Figures 4b**–**4f**). Notably, nitrogen (N) vacancies result in a greater band gap reduction compared to carbon (C) vacancies, as observed in **Figure 6b**. The band gap decreases further in structures with double vacancies. In the case of g-$C_3N_4$ with a double N vacancy (**Figure 4e**), the band gap nearly vanishes, rendering the defect structure nearly metallic. The TDOS plots also reveal the presence of a magnetic moment whenever asymmetry occurs between spin-up and spin-down states. Simulation results (**Table S2**, **Figure S9**) indicate that a magnetic moment emerges in structures with a single N vacancy or a combined C+N vacancy. However, symmetry is restored in all other defective structures.

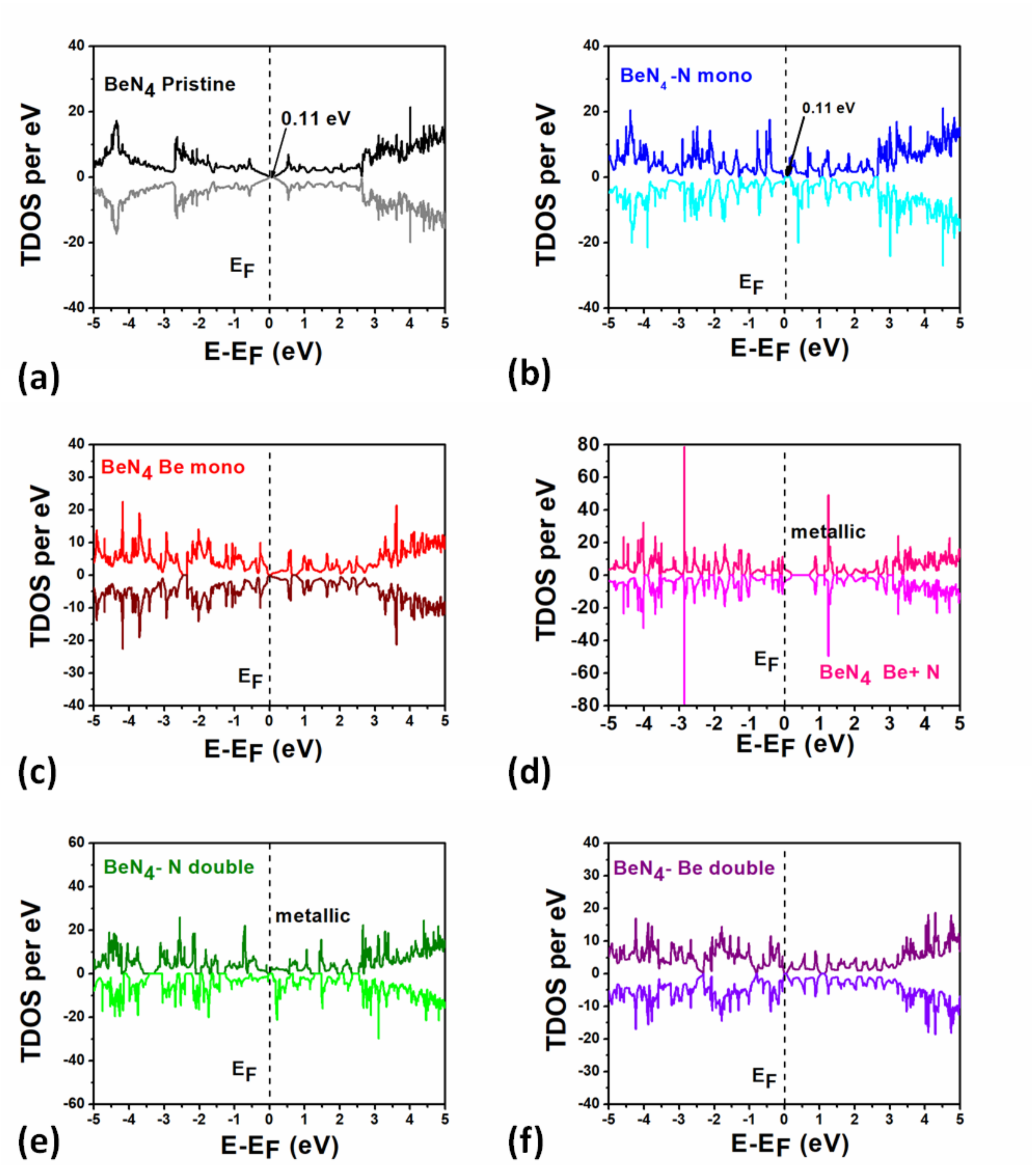

**Figure 5 :** Total density of states (TDOS) plots for pristine and defective beryllonitrene ($BeN_4$) structures.(a) Pristine $BeN_4$ (solid black: spin-up, solid grey: spin-down),(b) $BeN_4$ with nitrogen monovacancy (solid blue: spin-up, solid cyan: spin-down),(c) $BeN_4$ with berrylium monovacancy (solid red: spin-up, solid brown: spin-down),(d) $BeN_4$ with Be + nitrogen vacancy (solid pink: spin-up, solid magenta: spin-down),(e) $BeN_4$ with nitrogen double vacancy (solid dark green: spin-up, solid light green: spin-down),(f) $BeN_4$ with berrylium double vacancy (solid purple: spin-up, solid violet: spin-down).

Figure 5 presents the TDOS plots for both pristine and defective structures of beryllonitine. The pristine $BeN_4$ exhibits a minimal band gap of 0.11 eV. While the band gap remains unchanged for nitrogen monovacancy- defective structure, the density of states near the Fermi level increases compared to the pristine structure, as shown in Figure 5b. Other types of vacancies (Figures 5d, 5e) significantly reduce the band gap **(**Figures 5c, 5f) or even render the material metallic. A magnetic moment arises when asymmetric spin-up and spin-down states are introduced. Simulations indicate that magnetic moments occur exclusively in $BeN_4$ structures with nitrogen vacancy defect.

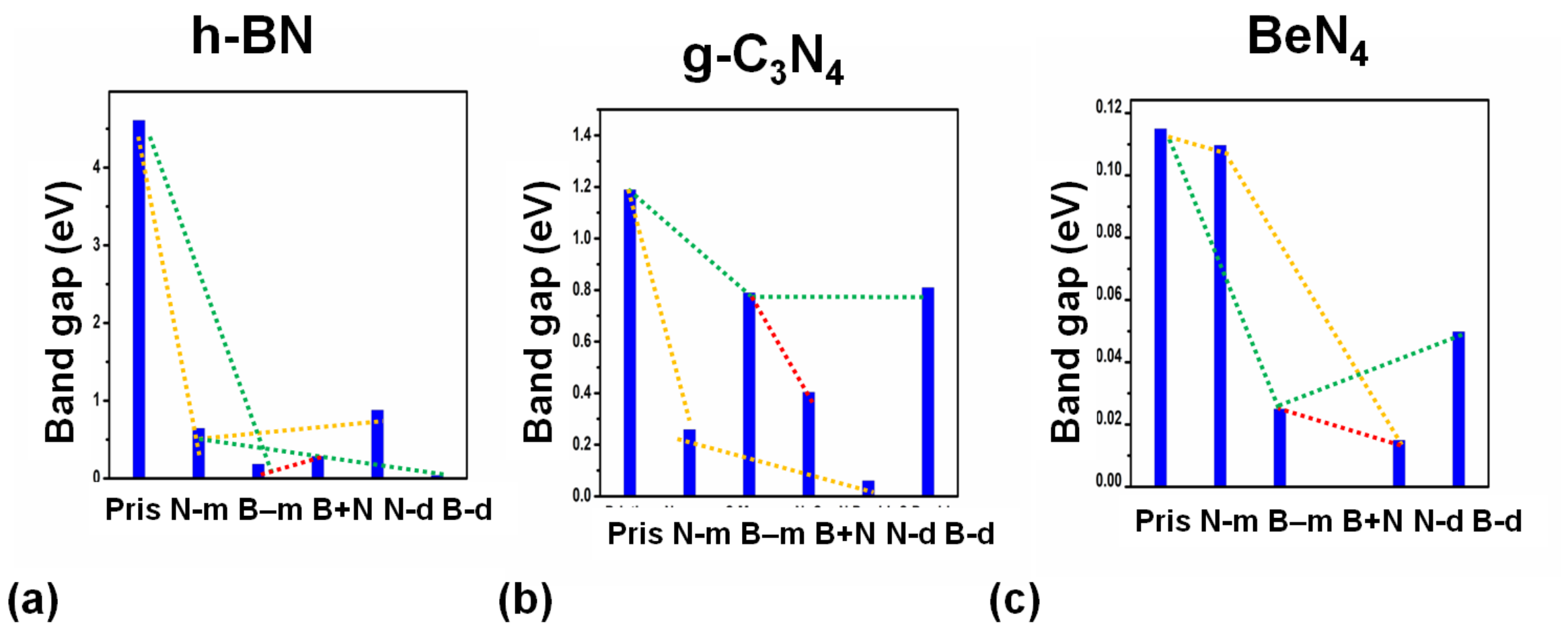

**Figure 6**: Modulation of Band gap values. From left to right; $M_xN_y$ Pristine; $M_xN_y$with a N vacancy; $M_xN_y$ with a M atom vacancy; $M_xN_y$ with a N and M atom double vacancy; $M_xN_y$with a N double vacancy; $M_xN_y$ with a M double vacancy where M stands for (a) boron in h-BN(b) carbon in g-$C_3N_4$ and (c) beryllium in $BeN_4$ respectively. Green dotted line indicates the change in band gap due to M type of vacancy whereas yellow line indicates change in band gap due to N vacancy.

### *3.5 Band Structure Plot*

For further analysis, the band structures of both pristine and defective structures were generated and are presented in Figures 7 to 9. Figures 7a–7f illustrate the band structures of pristine and vacancy-defective h-BN. The high-symmetry points Γ → M → K → Γ were selected for h-BN. Figure 7a displays the band structure of pristine h-BN, revealing a large indirect band gap of 4.6 eV, which aligns with the band gap value obtained from the TDOS plot and reported data from the literature. In the pristine structure, the conduction band minimum is located at the Γ point, while the valence band maximum is at K, confirming the indirect band gap nature of the pristine structure. Additionally, the pristine structure exhibits a net magnetic moment of zero, resulting in the overlap of spin-up and spin-down bands.

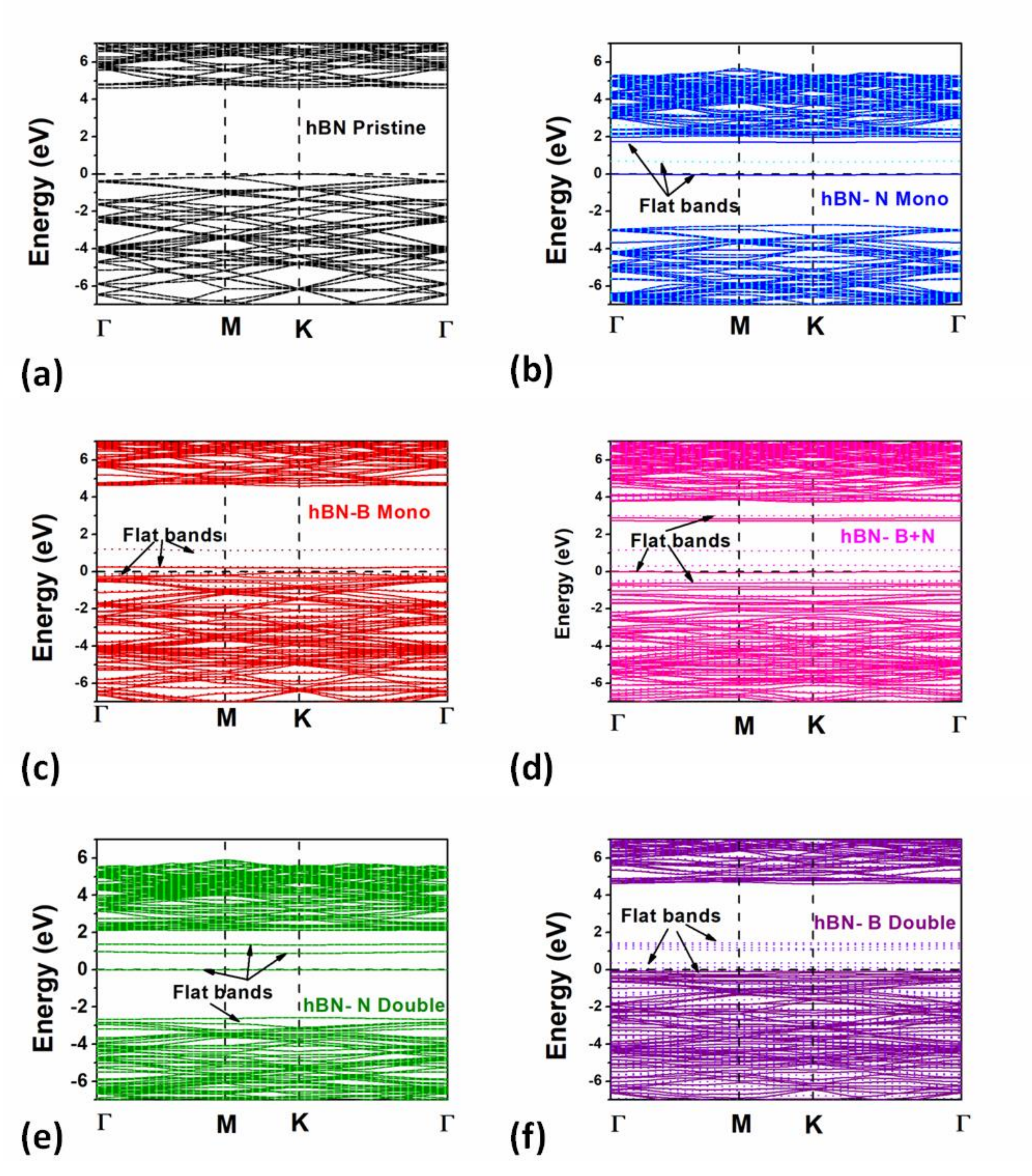

**Figure 7:** Band structure plots of hexagonal boron nitride (h-BN): (a) pristine h-BN (solid black line: spin-up channel; solid grey dotted line: spin-down channel), (b) h-BN with nitrogen monovacancy defect (solid blue line: spin-up; solid cyan dotted line: spin-down), (c) h-BN with boron monovacancy defect (solid red line: spin-up; solid brown dotted line: spin-down), (d) h-BN with boron + nitrogen vacancy defect (solid pink line: spin-up; solid magenta dotted line: spin-down), (e) h-BN with nitrogen double vacancy defect (solid dark green line: spin-up; solid light green dotted line: spin-down), (f) h-BN with boron double vacancy defect (solid purple line: spin-up; solid violet dotted line: spin-down).

**Figure 7b** presents the band structure of h-BN with an N vacancy defect, where the appearance of new LUMO states within the band gap is clearly visible. These states elevate the Fermi level from −5.2 eV to −2.68 eV, ultimately reducing the band gap to 0.64 eV. Similarly, new LUMO states are observed across all defect cases (**Figures 7c–7f**). The band gap values obtained from band structure analysis align with the TDOS results listed in **Table S2**. For N and B monovacancies, the band gap remains indirect, with the valence band maximum (VBM) at the Γ symmetry point and the conduction band minimum (CBM) at K. In the B vacancy structure, the band gap is further reduced, with the VBM shifting to K and the CBM located between Γ and K. **Figure 7d** highlights the transition to a direct band gap in defective h-BN structures. **Figure 7f** illustrates the band structure of a boron double-vacancy defect in h-BN, revealing the presence of spin-down bands at the Fermi level, indicative of its half-metallic nature. This finding aligns with the TDOS plots of the structure. The half-metallic property of these defects makes them highly valuable for applications in spintronics[65], magnetism, and quantum computing.

Another key observation is the emergence of flat bands in all defective structures. The flat bands observed in the band structure arise due to the introduction of localized electronic states associated with vacancy defects. These defects disrupt the periodic potential of the crystal, leading to electronic states that are spatially confined and energetically localized, which appear as flat bands near the Fermi level. The presence of such flat bands indicates a high density of states in a narrow energy range, which can significantly influence the electronic and transport properties of the material. In particular, these defect-induced states can act as trap sites, modify carrier lifetimes, and enhance field emission by reducing the effective work function. Spin-polarized flat bands near the Fermi level appear in each case, indicating strong electron-electron interactions that can give rise to exotic quantum phenomena such as superconductivity[66] and the anomalous Hall effect[67]. Flat bands are generally challenging to engineer, and having them spin-polarized is critically important for developing unconventional electronic devices that exploit spin degrees of freedom rather than charge, enabling multiple new functionalities. The systematic creation of vacancy defects in 2D nitride structures can facilitate flat-band engineering, which is essential for next-generation applications. The spin-polarized nature of these flat bands makes them even more compelling, as they combine the benefits of strong correlation effects with spin polarization, enabling more exotic and tunable quantum phases for spintronics applications[68,69].

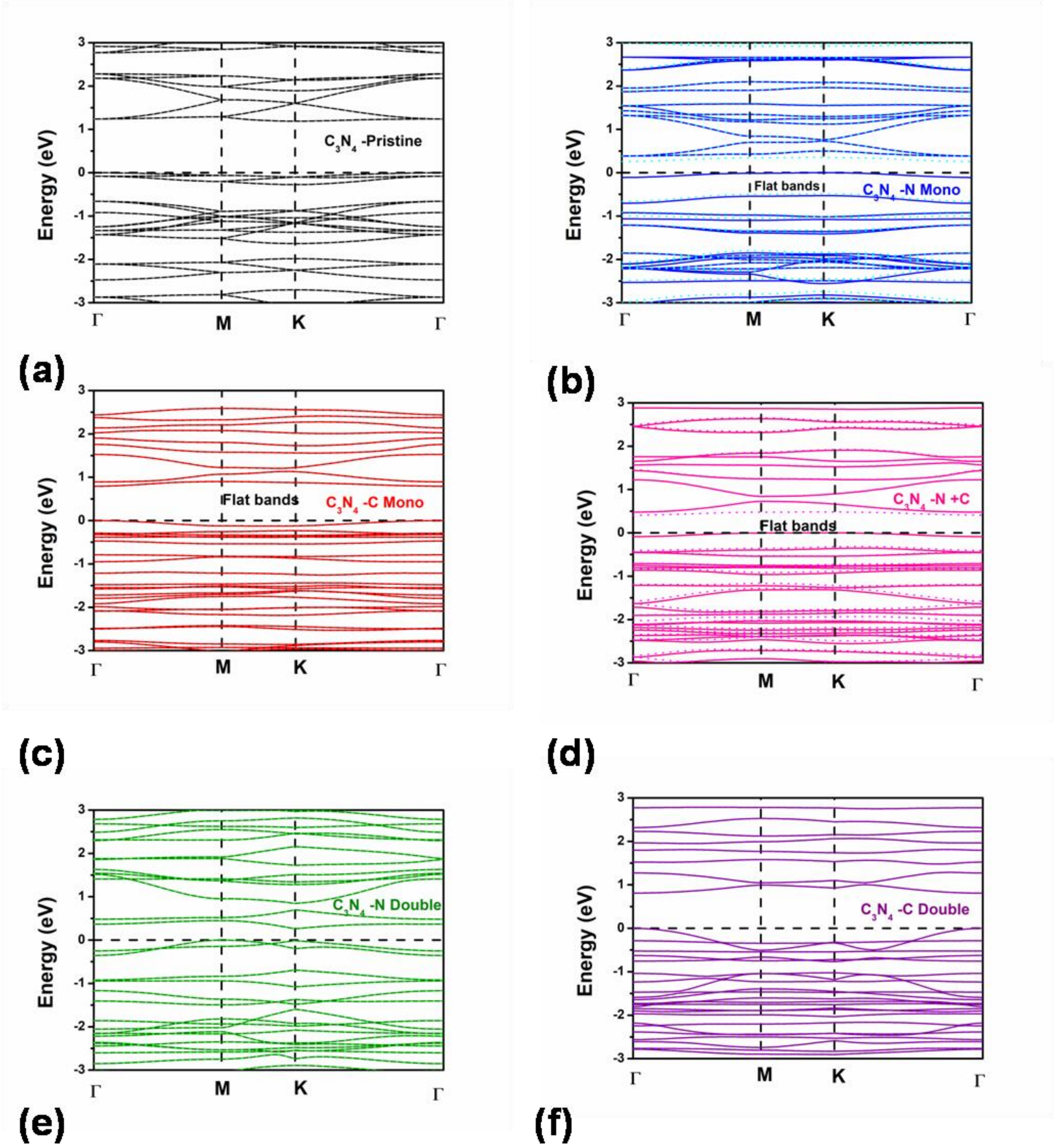

**Figure 8** : Band structure plots of graphitic carbon nitride (g-$C_3N_4$): (a) Pristine g-$C_3N_4$ (solid black line: spin-up channel; solid grey dotted line: spin-down channel), (b) g-$C_3N_4$ with nitrogen

monovacancy defect (solid blue line: spin-up; solid cyan dotted line: spin-down), (c) g-$C_3N_4$ with carbon monovacancy defect (solid red line: spin-up; solid brown dotted line: spin-down), (d) g-$C_3N_4$ with combined carbon and nitrogen vacancy defects (solid pink line: spin-up; solid magenta dotted line: spin-down), (e) g-$C_3N_4$ with nitrogen double vacancy defect (solid dark green line: spin-up; solid light green dotted line: spin-down), (f) g-$C_3N_4$ with carbon double vacancy defect (solid purple line: spin-up; solid violet dotted line: spin-down)

For g-$C_3N_4$, the high-symmetry points $\Gamma \rightarrow M \rightarrow K \rightarrow \Gamma$ were selected for band structure analysis. **Figure 8a** presents the band structure of the pristine structure, revealing a band gap of 1.19 eV, which aligns with the findings from the TDOS plot and the reported literature. A similar trend is observed in defective structures, where the creation of LUMO states due to vacancy defects raises the Fermi energy, ultimately reducing the band gap. In the case of carbon vacancy structures (**Figures 8c** and **8f**), a direct band gap is observed at the Γ high-symmetry point. Conversely, for nitrogen vacancies, the band gap remains indirect. **Figures 8b** to **8d** highlight the presence of flat bands near the Fermi level. Additionally, spin polarization-induced degeneracy is observed in defect structures with nitrogen mono vacancy only. **Figure 8** depicts the emergence of defect bands within the band gap region due to vacancy defects. The introduction of such vacancy defects in 2D structures typically gives rise to mid-gap states that significantly influence both optical transitions and carrier dynamics. These mid-gap states allow new sub-band gap optical transitions by enabling electron excitation from the valence band to the defect states or from the defect states to the conduction band. Consequently, the material exhibits enhanced optical absorption at lower photon energies, resulting in a red shift in the absorption spectrum. This behaviour can enhance the performance of photodetectors and photovoltaic devices by broadening their optical response.

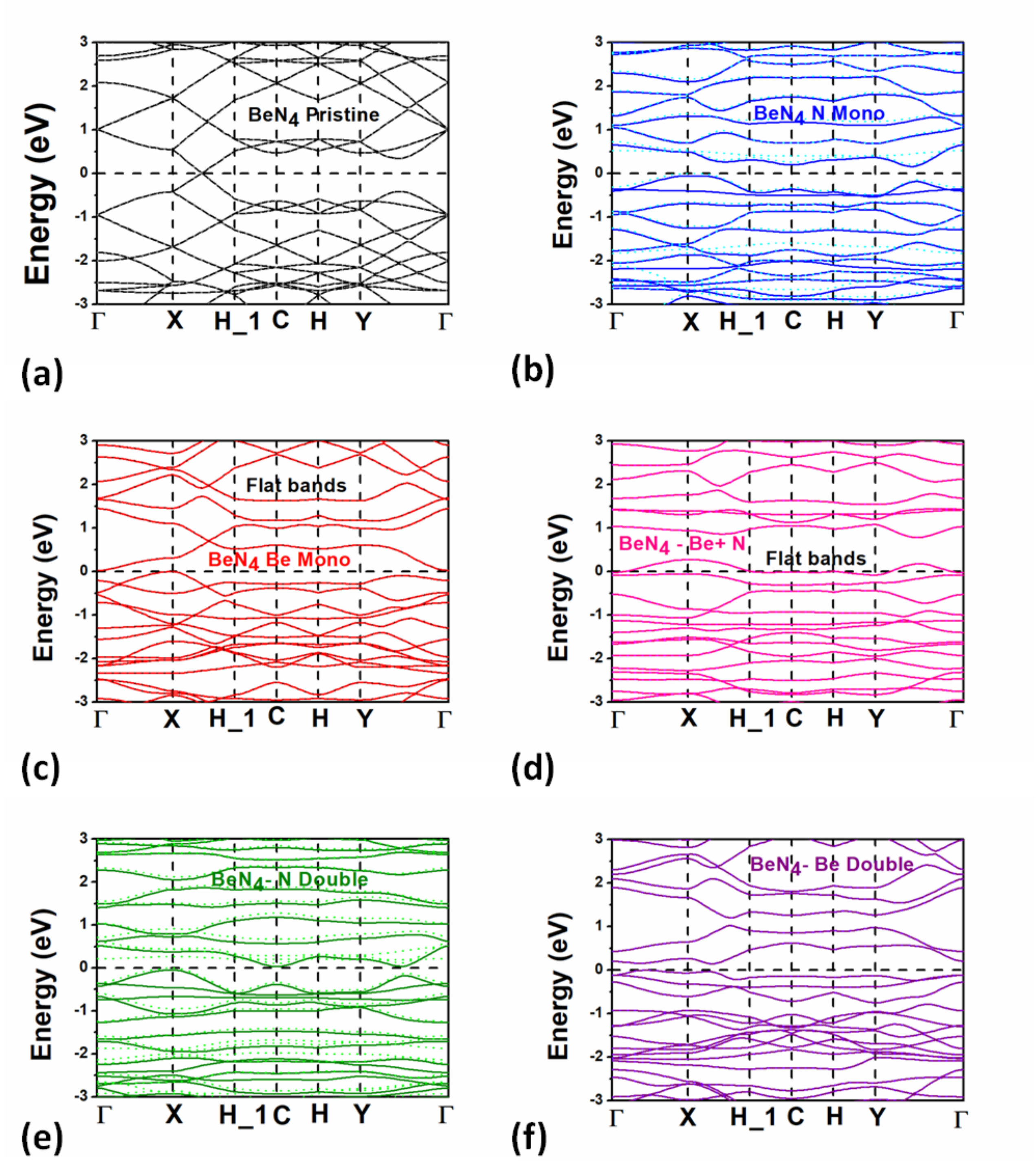

**Figure 9.** Band structure plots of Beryllonitrene ($BeN_4$): (a) Pristine $BeN_4$ (solid black line: spin up channel; solid grey dotted line: spin-down channel), (b) $BeN_4$ with nitrogen monovacancy defect (solid blue line: spin-up; solid cyan dotted line: spin-down), (c) $BeN_4$ with beryllium monovacancy defect (solid red line: spin-up; solid brown dotted line: spin-down), (d) $BeN_4$ with combined beryllium and nitrogen vacancy defects (solid pink line: spin-up; solid magenta line: spin-down), (e) $BeN_4$ with nitrogen double vacancy defect (solid dark green line:

spin-up; solid light green dotted line: spin-down), (f) $BeN_4$ with beryllium double vacancy defect (solid purple line: spin-up; solid violet dotted line: spin-down).

For $BeN_4$, the high-symmetry points $\Gamma \rightarrow X \rightarrow H_1 \rightarrow C \rightarrow H \rightarrow Y \rightarrow \Gamma$ were selected for band structure analysis. **Figure 9a** highlights the presence of an anisotropic Dirac cone at the Fermi level, though not at a high-symmetry point. **Figures 9b–9f** display the band structures of defective $BeN_4$, which do not exhibit the anisotropic Dirac cone. However, the defective structures show either a negligible band gap or metallic behaviour. Flat bands are relatively uncommon in $BeN_4$, but in Be mono-vacancy defect structures, the flat bands appear between the $H_1 \rightarrow C \rightarrow H$ symmetry points. The band structure also reveals spin polarization-induced degeneracy, occurring exclusively in nitrogen vacancy defect structures, leading to a net magnetic moment.

# 4. Discussion

In order to fully understand the changes occurring in the structures after the creation of defects, Bader charge analysis and partial density of states of the pristine and defective structures were calculated. Bader charge analysis quantifies the charge distribution in 2D materials by partitioning electron density and assigning charge to each atom. Table S4 to S6 in the supporting information tabulate the Bader charge values on each individual atom in the pristine and defective structure of the materials under investigation. For the analysis in this work, we have calculated the difference in charge in the pristine ($Q_{pristine}$) and defective structures ($Q_{defect}$) of each atom following **Equation 3.**

$$\Delta = Q_{defect} - Q_{pristine} \tag{3}$$

Bader transfer plot of h-BN with an N monovacancy is depicted in **Figure 10** which shows the highest reduction in work function from 5.97 eV (pristine) to 3.45 eV (vacancy defect) among all the vacancy defects. The Bader transfer plot shown in **Figure 10a** shows a gain in charge on the nearest boron atom as well as the neighbouring nitrogen atoms to the defect site after the removal of a single nitrogen atom. To accommodate this excess charge, new LUMO states are formed owing to the existence of dangling bonds in the region (**Figure 10a**). These new electronic levels just below the conduction band result in the increase in Fermi energy (**Table 2**) which in turn leads to lowering of the work function. In order to ascertain that these new levels are introduced mainly because of the defect and its neighbouring atoms, PDOS plots of the neighbouring atoms adjacent to the defect site have been carried out and the simulation results are shown in **Figures 10b** to **10e**. The PDOS plots reveal that the new states at the Fermi energy level are contributed primarily by the nearest Boron atom followed by the nitrogen atoms (Atom number 40, 48) which experience charge gain. In the background of these plots, TDOS of the entire structure is shown as a grey line. From the figure it is evident that neighbouring atoms to the defect site are contributing to the change of the total density of states. As a comparison an atom (47) placed far from the defect site does not experience charge gain and has no contribution to the density of states at the Fermi energy level which is akin to the pristine structure. The average charge on the nitrogen atoms increases from 8 in pristine to 8.106 in the nitrogen vacancy defect.

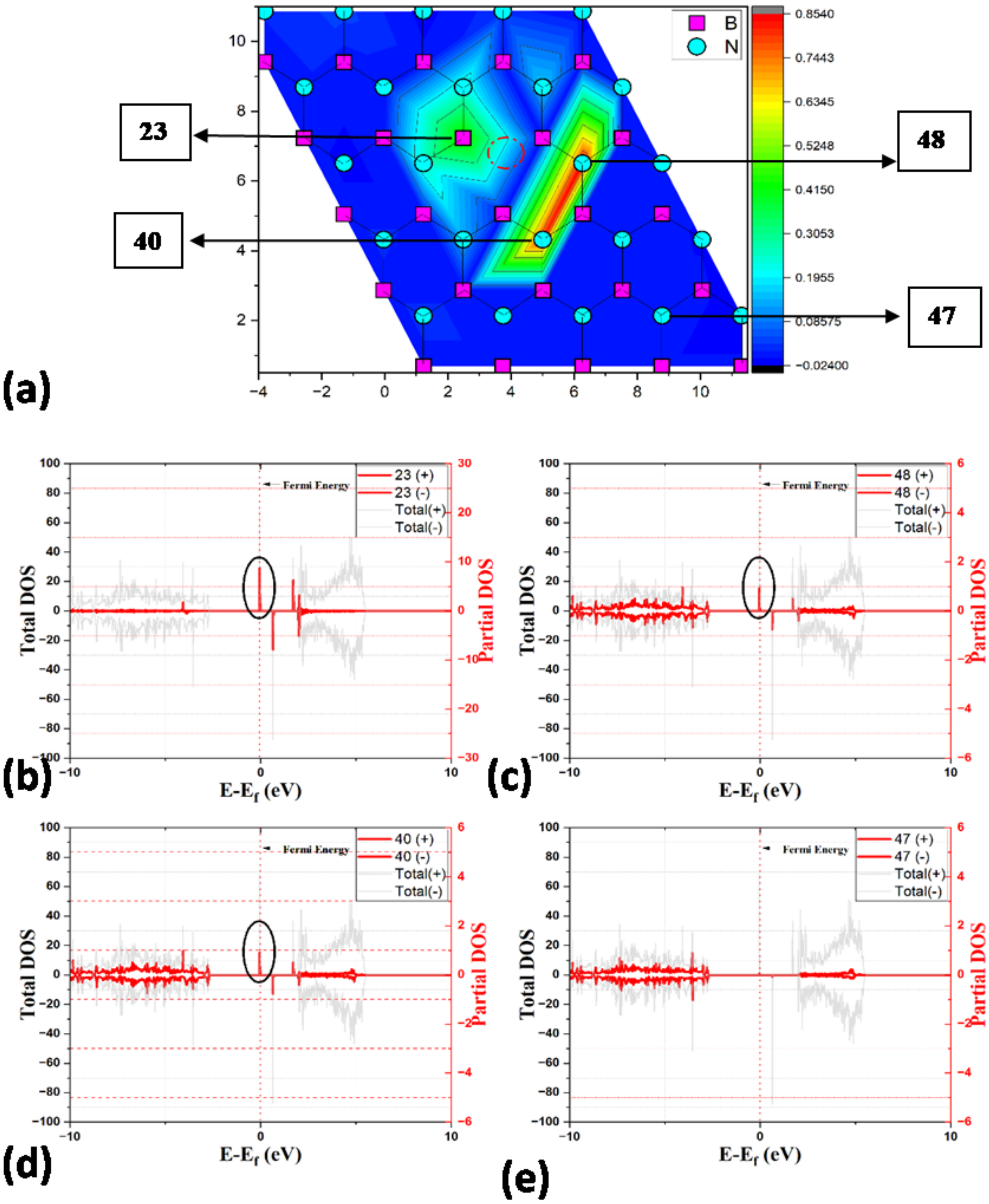

**Figure 10**. Analysis plots for h-BN with a nitrogen mono-vacancy defect: (a) Bader charge difference plot for the N mono-vacancy defect. Magenta squares represent boron atoms, and cyan circles represent nitrogen atoms in the supercell. The defect site is marked with a red circle. The plot shows net charge gain on the neighboring boron and nitrogen atoms (highlighted in light green and red regions). Atoms numbered 23, 40, 48, and 47 are indicated with black arrows. (b) PDOS plot for atom 23 (nearest boron atom to the defect site), with the x axis aligned to the Fermi energy. Note that the scales for PDOS (right) and TDOS (left) differ. (c) PDOS plot for atom 48 (nearest nitrogen atom to the defect site exhibiting charge gain), with the x-axis aligned to the Fermi energy. Scales for PDOS (right) and TDOS (left) differ. (d) PDOS plot for atom 40 (nitrogen atom near the defect site exhibiting charge gain), with the x axis aligned to the Fermi energy. Scales for PDOS (right) and TDOS (left) differ. (e) PDOS plot for atom 47 (nitrogen atom distant from the defect site), with the x-axis aligned to the Fermi energy. Scales for PDOS (right) and TDOS (left) differ. Black ellipses highlight states at the Fermi energy level in the partial DOS contributions for the respective atoms.

**Figure S10** presents the analysis of a nitrogen mono vacancy defect in g-$C_3N_4$, which exhibits the lowest work function (4.3 eV) among all vacancy defects of graphitic carbon nitride. The Bader charge transfer plot in **Figure S10a** reveals intriguing trends similar to those observed in h-BN. In g-$C_3N_4$, localized charge density increases around atoms neighbouring the defect site, leading to a reduction in the work function. However, the charge accumulation is more pronounced at specific nitrogen atoms at the bridgehead sites rather than near the vacancy itself. Bridgehead nitrogen atoms, which are $sp^2$-hybridized and connected to three heptazine units, play a crucial role in this behaviour. These heptazine rings, being aromatic with 14 π-electrons, facilitate electron delocalization. Upon vacancy formation, excess charge delocalizes and concentrates at bridgehead nitrogen sites, as shown in **Figure S10a**. However, PDOS plots (**Figures S10b–S10e**) indicate that these nitrogen atoms do not contribute to the density of states at the Fermi level. This can be attributed to their initial charge (6.68) in the pristine structure, which allows them to accommodate unpaired electrons in the defective structure, raising their charge to 7.78 without introducing new states. Instead, the new LUMO states near the conduction band originate from neighbouring carbon and nitrogen atoms (Atoms 23 and 35). This results in an increase in Fermi energy from -4.31 eV (pristine) to -3.55 eV (with nitrogen vacancy), thereby decreasing the work function, as summarized in **Table 2.**

**Figure S11** presents simulation results illustrating structural changes in $BeN_4$ upon the creation of a nitrogen vacancy defect. The Bader charge transfer plot **(Figure S11a**) shows charge accumulation in neighbouring nitrogen atoms (red and yellow regions) after the mono-vacancy is introduced. Notably, $BeN_4$ exhibits the smallest charge gain ($\Delta$ = 0.2) among the investigated nitride samples. The simulations suggest that charge redistribution in 2D nitride structures ($M_xN_\gamma$, where M = B/ C/ Be) depends on both the nitrogen content and the nature of the M–N bond. In $BeN_4$, the Be:N ratio is 1:4 (compared to 3:4 in g-$C_3N_4$ and 4:4 in h-BN), resulting in an average charge of 5 in the pristine structure. When a nitrogen vacancy forms, the redistributed charge is absorbed by neighbouring nitrogen atoms acting as charge sinks. Additionally, electron loss due to the missing nitrogen leads to partial charge depletion in adjacent nitrogen atoms (blue regions). **Figures S11b–S11e** display the PDOS plots for labelled atoms, revealing that the nitrogen atom with charge gain develops new states at the Fermi energy level. Overall, the combined effects of charge loss and gain lead to a slight decrease in Fermi energy and an increase in work function (**Table 2**). The simulations further indicate that the magnitude of the work function change in nitrides is influenced by the polarity of the nitrogen bond with the other element. In h-BN, the B–N bond is more polar covalent compared to C–N in g-$C_3N_4$, resulting in stronger localized electronic state disturbances upon vacancy formation. Consequently, h-BN experiences a more significant work function shift than g-$C_3N_4$. These findings highlight that nitrogen vacancy defects can effectively tune the work function of 2D nitrides, making them promising for various electronic applications. Our simulation also indicates that defect engineering in 2D nitride structures can be useful for tuning the band gap of the material where the band gap values depend on the type of vacancy defect and their density.

Bader charge analysis of various vacancy defects in h-BN, g-$C_3N_4$, and $BeN_4$ is shown in **Figures S12–S14**. For boron and carbon mono vacancies (**Figures S12a** and **S13a**), as well as their respective double vacancies (**Figures S12d** and **S13d**), a significant charge loss is observed near the defect sites, particularly on neighbouring nitrogen atoms. This electron depletion leads to a reduction in the Fermi energy and an increase in the work function due to the formation of charge sinks that hinder electron emission from the surface. In contrast, nitrogen mono-vacancies and double vacancies in both h-BN and g-$C_3N_4$ (**Figures S12c,** and **S13c**) exhibit charge accumulation near the defect sites. This results in the emergence of new states near the conduction band, raising the Fermi energy and thereby lowering the work function. For mixed boron/carbon + nitrogen double vacancies (**Figures S12b** and **S13b**), both charge loss and gain occur simultaneously, leading to a net minimal change in Fermi energy and work function due to the competing effects of both elemental vacancies. **Figure S14** presents the charge transfer profiles for $BeN_4$ vacancy defects. Vacancies involving Be atoms (monovacancy, Be + N double, and Be double vacancies) show substantial charge loss, attributed to the limited number of Be donor atoms, leading to reduced Fermi energy and increased work function. On the other hand, nitrogen double vacancies (**Figure S14c**) result in localized charge accumulation, consistent with trends seen in h-BN and g-$C_3N_4$, and account for the minimal increase in work function in these structures.

## 5. Conclusion

Our comprehensive simulation study on 2D nitride materials namely hexagonal boron nitride (h-BN), graphitic carbon nitride (g-$C_3N_4$), and beryllonitrene ($BeN_4$) offers critical insights into the role of vacancy defects in tailoring electronic properties for device-level applications. The introduction of a nitrogen monovacancy in h-BN results in a remarkable 1.53 eV (-42%) reduction in work function, the most significant among the studied materials, primarily due to charge redistribution and the formation of localized states near the Fermi level, as confirmed through Bader charge analysis and partial density of states (PDOS). In g-$C_3N_4$, nitrogen vacancies lead to concurrent reductions in both work function and band gap, particularly in double vacancy configurations, which show a tendency toward metallic behaviour. $BeN_4$, on the other hand, exhibits relatively stable electronic characteristics upon defect introduction, attributed to its distinct bonding environment and lower nitrogen concentration. Band structure analysis corroborates these findings, revealing defect-induced flat bands near the Fermi level in h-BN and g-$C_3N_4$, with the double boron-vacancy in h-BN notably exhibiting half-metallic behaviour underscoring its promise for spintronic applications. These defect-induced modulations also have significant implications for optoelectronic device engineering. Reduced band gaps and the presence of mid-gap states can enable sub-band gap photon absorption, potentially extending the spectral response range of photodetectors, light emitters, and solar cells based on 2D nitrides. Ultimately, our findings establish a theoretical foundation demonstrating the efficacy of nitrogen vacancy defects in modulating the electronic properties of 2D nitride materials, paving the way for their tailored application in next-generation optoelectronic and spintronic devices.

## 6. Acknowledgement

We extend our gratitude to the BARC supercomputing facilities and the dedicated staff members for their invaluable support.

## 7. Contributions

S.G.S performed all the DFT calculations. S.G.S and K.P.S analysed the data and wrote the initial draft of the paper. B.C supervised the overall work, contributed to the analysis, and provided revisions to the paper. All authors discussed the results and contributed to the final version of the paper.

## 8. Competing Interests

The authors declare that they have no competing interests.

## 9. Data Availability

The authors confirm that the main data supporting the findings of this study are available within the paper and its supplementary materials. Furthermore, other relevant data can be made available from the corresponding author upon reasonable request.

## 10. Supporting Information

The Supporting Information is available.

## 11. References

(1) Novoselov, K. S.; Geim, A. K.; Morozov, S. V.; Jiang, D. E.; Zhang, Y.; Dubonos, S. V.; Grigorieva, I. V.; Firsov, A. A. Electric field effect in atomically thin carbon films. *Science* **2004**, *306*, 666−669.

(2) Chhowalla, M.; Shin, H. S.; Eda, G.; Li, L. J.; Loh, K. P.; Zhang, H. The chemistry of two dimensional layered transition metal dichalcogenide nanosheets. *Nat. Chem.* **2013**, 263−275.

(3) Li, L.; Yu, Y.; Ye, G. J.; Ge, Q.; Ou, X.; Wu, H.; Feng, D.; Chen, X. H.; Zhang, Y. Black Phosphorus Field-Effect Transistors. *Nat. Nanotechnol.* **2014**, *9*, 372−377.

(4) Dean, C. R.; Young, A. F.; Meric, I.; Lee, C.; Wang, L.; Sorgenfrei, S.; Watanabe, K.; Taniguchi, T.; Kim, P.; Shepard, K. L.; Hone, J. Boron nitride substrates for high-quality graphene electronics. *Nat. Nanotechnol.* **2010**, *5*, 722−726.

(5) Castro Neto A. H. ; Guinea F.; Peres N. M. R.; Novoselov K. S.; Geim A. K. The electronic properties of graphene, *Rev. Mod. Phys* **2009**, 81, 10 .

(6) Wang, Q. H.; Kalantar-Zadeh, K.; Kis, A.; Coleman, J. N.; Strano, M. S. Electronics and optoelectronics of two-dimensional transition metal dichalcogenides. *Nat. Nanotechnol.* **2012**, 699−712.

(7) Akinwande, D.; Petrone, N.; Hone, J. Two-dimensional flexible nanoelectronics. *Nat Commun*. **2014** 5, 5678.

(8) Xu H. Han : X. ; Liu W.; Liu P. ; Fang H.; Li X.; Li Z. e.t al. Ambipolar and Robust WSe2 Field-Effect Transistors Utilizing Self-Assembled Edge Oxides. *Adv. Mater. Interfaces* **2020**, 7, 1901628.

(9) Novoselov, K. S.; Jiang, Z.; Zhang, Y.; Morozov, S. V.; Stormer, H. L.; Zeitler, U.; Maan, J. C.; Boebinger, G. S.; Kim, P.; Geim, A. K. Room-Temperature Quantum Hall Effect in Graphene. *Science* **2007**, *315* (5817), 1379–1379.

(10) Bolotin, K. I.; Sikes, K. J.; Jiang, Z.; Klima, M.; Fudenberg, G.; Hone, J.; Kim, P. ; Stormer, H. L. Ultrahigh electron mobility in suspended graphene *Solid State Commun*. **2008**, 146, 3517,

(11) Pham P. V.; Bodepudi S.C.; Shehzad K.; Liu Y.Xu Y.; Yu B.; Duan X. 2D Heterostructures for Ubiquitous Electronics and Optoelectronics: Principles, Opportunities, and Challenges *Chem. Rev.* **2022** 122,6,6514-6613.

(12) Liu, M.; Yin, X. B.; Ulin-Avila, E.; Geng, B. S.; Zentgraf, T.; Lu, L.; Wang, F.; Zhang, X. A graphene-based broadband optical modulator *Nature* **2011**, 474, 645.

(13) Zhu, Y. W.; Murali, S.; Stoller, M. D.; Ganesh, K. J.; Cai, W. W.; Ferreira, P. J.; Pirkle, A.; Wallace, R. M.; Cychosz, K. A.; Thommes, M.; Su, D.; Stach, E. A.; Ruoff, R. S. Carbon-Based Supercapacitors Produced by Activation of Graphene *Science* **2011**15377, 332.

(14) Deng, M.; Yang, X.; Silke, M.; Qiu, W. M.; Xu, M. S.; Borghs, G.; Chen, H. Z. Electrochemical deposition of polypyrrole/graphene oxide composite on microelectrodes towards tuning the electrochemical properties of neural probes Se*nsors Actuators B: Chem.***2011**, 158, 17611.

(15) Han, W., Kawakami, R. K., Gmitra, M., & Fabian, J. Graphene spintronics. *Nature nanotechnology* **2014** *9*(10), 794-807.

(16) Molaei M. J.; Younas M.; Rezakazemi M. A Comprehensive Review on Recent Advances in Two-Dimensional (2D) Hexagonal Boron Nitride *ACS Appl. Electon. Mater.***2021** 3 (12), 5165-5187.

(17) Shen, Y.; Zhu, K., Xiao; Y. et al. Two-dimensional-materials-based transistors using hexagonal boron nitride dielectrics and metal gate electrodes with high cohesive energy. *Nat Electron* **2024**7, 856–867.

(18) Hui F.; Pan C.;Shi Y. ; Ji Y. ;Grustan-Gutierrez E.; Lanza M. On the use of two dimensional hexagonal boron nitride as dielectric, *Microelectron. Eng.* **2016** 163, 119-133,

(19) Kroke E.; Schwarz M.; Horath-Bordon E.; Kroll P.; Noll B.; Norman AD Tri-s-triazine derivatives. Part I. From trichloro-tri-s-triazine to graphitic $C_3N_4$ structures. *New J Chem* **2002**26(5):508–512.

(20) Chen X.; Liu Q.; Wu Q.; Du P.; Zhu J.; Dai S.; Yang S. Incorporating graphitic carbon nitride (g-C3N4) quantum dots into bulk-heterojunction polymer solar cells leads to efficiency enhancement. *Adv. Funct. Mater.***2016** 26 1719–1728.

(21) Ragupathi V.; Panigrahi P.; Subramaniam N. G. Band gap engineering in graphitic carbon nitride: Effect of precursors *Optik***2020** 202,163601.

(22) Jiang L.; Yuan X. ; Pan Y. ; Liang J. ; Zeng G. ; Wu Z. ; Wang H. Doping of graphitic carbon nitride for photocatalysis: A review *Appl. Catal. B Environ.* **2017** 217, 388-406.

(23) Cao S.; Low J.; Yu J.;Jaroniec M. Polymeric photocatalysts based on graphitic carbon nitride. *Adv Mater.* **2015** 27(13) ,2150–2176.

(24) Adegoke K.A. ; Maxakato N.W.; Efficient strategies for boosting the performance of 2D graphitic carbon nitride nanomaterials during photoreduction of carbon dioxide to energy-rich chemicals *Mater. Today Chem.* **2022** 23 100605.

(25) Thomas A.A.; Pallavolu M.R. ; , Khan M.E.; Cherusseri J. Graphitic carbon nitride (g-C3N4): Futuristic material for rechargeable batteries, *J. Energy Storage*, **2023** 68, 107673.

(26) Kamble B.B.; Sharma K.K.; Sonawane K.D.;Tayade S.N.; Grammatikos S.; Reddy Y.V.M.; Reddy S.L.; Shin J. H.; Park J.P. Graphitic carbon nitride-based electrochemical sensors: A comprehensive review of their synthesis, characterization, and applications *Adv. Colloid Interface Sci.* **2023** 333,103284.

(27) Liao G.; He F.; Qing Li; Zhong L.; Zhao R.; Che H.; Gao H.; Fang B. Emerging graphitic carbon nitride-based materials for biomedical applications *Prog. Mater, Sci***2020** 112, 100666.

(28) Maxim Bykov M.; Fedotenko T.; Chariton S.; Laniel D..; Glazyrin K.; Hanfland M.; et.al High-Pressure Synthesis of Dirac Materials: Layered van der Waals Bonded $BeN_4$ Polymorph *Phys. Rev. Lett.* 2021 126, 175501.

(29) Bafekry A.; Stampfl C.; Faraji M.; Yagmurcukardes M.; Fadlallah M. M.; Jappor H. R.; Ghergherehchi M.; Feghhi S. A. H. A dirac-semimetal two-dimensional BeN4: thickness-dependent electronic and optical properties *Appl. Phys. Lett.*2021 118, 203103.

(30) Castro Neto, A. H.; Guinea, F.; Peres, N. M. R.; Novoselov, K. S.; Geim, A. K. The Electronic Properties of Graphene. *Rev. Mod. Phys.* **2009**, 81 (1), 109–162.

(31) Mortazavi B.; Shojaei F.; Zhuang X. Ultrahigh stiffness and anisotropic Dirac cones in $BeN_4$ and $MgN_4$ monolayers: A first-principles study *Mater. Today Nano.* **2021** *15*, 100125.

(32) Berdiyorov, G. R.; Mortazavi, B.; Hamoudi, H. Anisotropic Charge Transport in 1D and 2D $BeN_4$ and $MgN_4$ Nanomaterials: A First‐Principles Study. *FlatChem* **2022**, 31, 100327.

(33) Zhu M.; Li Q.; Zhang L.; Su J.; Yang C.; Wang, H. Electronic and optical properties of semiconducting $BeN_4$ nanoribbons, *J. Phys. Chem. Solids*2024 191, 112054.

(34) Shuyi Lin S.; Xu M.; Wang F.; Hao J.; Li Y. Ultrahigh energy density $BeN_4$ monolayer: A nodal-line semimetal anode for Li-ion batteries *Phys. Rev. Research.* 2024 6, 013028.

(35) Trivedi R.; Kaur S.; Garg N.; Chakraborty B. Ti-decorated nitrogen-rich $BeN_4$ monolayer for reversible hydrogen storage: DFT investigations, Applied Surface Science 2023 622,156806.

(36) Sanyal G.;T. Nair H.T.; Jha P.K.; Chakraborty B. First principles study on yttrium decorated $BeN_4$ monolayer for reversible hydrogen storage *J. Energy Storage.*2023 68, 107892.

(37) Ahmed, B.; Tahir, M. B.; Ali, A.; Sagir, M.; Nassani, A. A. First-Principles Study of the Structural and Electronic Properties of N-Doped $Zr_3C_2$ MXenes. *Mater. Sci. Semicond. Process.* **2025**, 192, 109417.

(38) Ahmed, B.; Tahir, M. B.; Ali, A.; Sagir, M. Exploring the Structural and Electronic Properties of N-Doped $Ti_2C$ MXenes for Novel Applications in Advanced Materials and Devices: A DFT Study. *Mater. Sci. Semicond. Process*. **2025**, 192, 109091.

(39) Kag D.; Luhadiya N.; Nagesh D.; Patil D.; Kundalwal S.I. Strain and defect engineering of graphene for hydrogen storage via atomistic modelling*Int. J. Hydrogen Energy* **2021** 46 (43) (2021) 22599-22610.

(40) Sarkar S. G.; Jethawa U.; Sanyal G.; Chakraborty B. Work Function Modulation in 2D Carbon Allotropes via Defect Engineering for Field Emission Applications: A DFT Analysis *ACS Appl. Electron. Mater*. **2024**, 6, 12, 8898–8911.

(41) Lin, S.; Ye, X.; Johnson, R. S.;Guo, H. First-principles investigations of metal (Cu, Ag, Au, Pt, Rh, Pd, Fe, Co, and Ir) doped hexagonal boron nitride nanosheets: stability and catalysis of CO oxidation. *J. Phys. Chem. C.***2013** *117*(33), 17319-17326.

(42) Chettri, B.; Patra P. K.; Vu, T. V.; Nguyen, C. Q.; Yaya, A.; Obodo, K. O.;Rai, D. P. et.al. Induced ferromagnetism in bilayer hexagonal Boron Nitride (h-BN) on vacancy defects at B and N sites. P*hysica E Low Dimens. Syst. Nanostruct.* **2021**126, 114436.

(43) Zhang, J.; Sun, R.; Ruan, D.; Zhang, M.; Li, Y.; Zhang, et.al. Point defects in two-dimensional hexagonal boron nitride: A perspective. *J. Appl. Phys* **2020** 128(10).

(44) Qu L. H.; Deng Z. Y.; Yu J.; Lu X. K.;Zhong C. G.; Zhou P. X.; et.al. Mechanical and electronic properties of graphitic carbon nitride (g-$C_3N_4$) under biaxial strain. *Vacuum* **2020** 176, 109358.

(45) Kumar A.; Raizada P.;Hosseini-Bandegharaei A.; Thakur V. K.; Nguyen V. H.; Singh P. C-, N-Vacancy defect engineered polymeric carbon nitride towards photocatalysis: viewpoints and challenges. *J. Mater. Chem. A* **2021** *9(1),* 111-153.

(46) Lakshmy S.; Sanyal G.; Kalarikkal N.; Chakraborty B. Influence of vacancy defects on 2D BeN4 monolayer for NH3 adsorption: a density functional theory investigation *Nanotechnology* **2023** 34 (43) 435504.

(47) Lakshmy S.; Banerjee A.; Sanyal G.; Kalarikkal N.; Chakraborty B. Influence of Be vacancy on 2D $BeN_4$ single-layer for enhanced $H_2S$ sensing: prediction from first-principles simulations *J. Phys. D: Appl. Phys.***2024** 57 , 275301

(48) Vamsi Krsihna B.; Ravi S.; Prakash M. D. Recent developments in graphene based field effect transistors, *Mater. Today Proc.* **2021** 45 (2) 1524-1528.

(49) DeSousa M.S.M; Fujun Liu F.; FanyaoQu F.; Chen W. Vacancy-engineered flat-band superconductivity in holey graphene, *Phys. Rev. B* **2022** 105.

(50) Hafner, J. Ab-initio simulations of materials using VASP: Density-functional theory and beyond *J. Comput. Chem.* **2008** 29(12), 2044–2078.

(51) Grimme, S. Semiempirical GGA-type density functional constructed with a long-range dispersion correction *J. Comput. Chem.* **2006** 27(17), 1787–1799.

(52) Wisesa, P.; McGill, K. A.; Mueller T. Efficient generation of generalized Monkhorst-Pack grids through the use of informatics. *Phys. Rev. B*, **2016** 93(15), 155109.

(53) Tang, W.; Sanville, E.; ṣHenkelman, G. A grid-based Bader analysis algorithm without lattice bias. *J. Phys. Condens. Matter* **2009** 21(8), 084204.

(54) Roy S.; et al. Structure, properties, and applications of two-dimensional hexagonal boron nitride *Adv. Mater.* **2021** *33*(44), 2101589.

(55) Topsakal M.; Aktürk E.; Ciraci S. J. P. R. B. First-principles study of two- and one-dimensional honeycomb structures of boron nitride *Phys. Rev. B* **2009** *79*(11), 115442.

(56) Amalia W.; Nurwantoro P. Density-functional-theory calculations of structural and electronic properties of vacancies in monolayer hexagonal boron nitride (h-BN) *Comput. Condens. Matter.* **2019** *18*, e00354.

(57) Mane P.;Vaidyanathan A.; Chakraborty B. Graphitic carbon nitride (g-$C_3N_4$) decorated with Yttrium as potential hydrogen storage material: Acumen from quantum simulations *Int. J. Hydrog. Energy.* **2022** 47 (99) , 41898-41910.

(58) Silva A. M.; Rojas M. I. Electric and structural properties of polymeric graphite carbon nitride (g-$C_3N_4$): A Density Functional Theory study *Comput. Theor. Chem.* **2016** *1098*, 41–49.

(59) Mortazavi B.; Shojaei F.; Zhuang X. Ultrahigh stiffness and anisotropic Dirac cones in $BeN_4$ and $MgN_4$ monolayers: A first-principles study *Mater. Today Nano.* **2021** *15*, 100125.

(60) Chakraborty, B.; Nandi, P. K.; Kawazoe, Y.; Ramaniah, L. M. Room-Temperature $d^0$ Ferromagnetism in Carbon-Doped $Y_2O_3$ for Spintronic Applications: A Density Functional Theory Study. *Phys. Rev. B* **2018**, *97* (18), 184411

(61) Dimov, N.; Staykov, A.; Kusdhany, I. M.; Lyth, S. M. Tailoring the Work Function of Graphene via Defects, Nitrogen-Doping and Hydrogenation: A First Principles Study. *Nanotechnology* **2023**, *34*, 415001.

(62) Jacobs R.; Morgan D.; Booske J. Work function and surface stability of tungsten-based thermionic electron emission cathodes. *APL Mater.* 2**017** 5 (11), 116105.

(63) Pakhira N.; Mahato R. Role of work function distribution on field emission effects, *Phys. B Condens. Matter.* **2023** 670,415394.

(64) Wang G.; Yang P.; Moody N.A.; Batista E.R. Overcoming the quantum efficiency-lifetime tradeoff of photocathodes by coating with atomically thin two-dimensional nanomaterials. *npj 2D Mater Appl*. **2018** 2, 17.

(65) de Groot R. A.; Mueller F. M.; Engen, P.; BuschowK. H. J. New class of materials: half-metallic ferromagnets. *Phys. Rev. Lett.***1983** 50, 2024–2027.

(66) Yankowitz M.; Chen S.; Polshyn H.; Zhang Y.; Watanabe K.; Taniguchi T.; Dean, C. R. Tuning superconductivity in twisted bilayer graphene. *Science* **2019** 363(6431), 1059-1064.

(67) Chang C. Z.; Liu C. X.; MacDonald A. H. Colloquium: Quantum anomalous hall effect *Rev. Mod. Phys***2023** 95(1), 011002.

(68) Jugovac M.; Iulia Cojocariu I.; Sánchez-Barriga J.; GargianiP.; Valvidares M.; Feyer V.; Blügel S.; Bihlmayer G.; Perna P. Inducing Single Spin-Polarized Flat Bands in Monolayer Graphene *Adv. Mater.***2023**, 35, 2301441.

(69) Bhattacharya A.; Timokhin I.; Chatterjee R.; Yang Q.; Mishchenko A. Deep learning approach to genome of two-dimensional materials with flat electronic bands. *Npj Comput. Mater.* **2023** 9(1), 101.

# Supporting Information

## Computational Insights into Defect Induced Modulation in Electronic Properties of 2D Nitride Monolayers

Shreya G Sarkar[1],Kuneh Parag Shah[2] and Brahmananda Chakraborty[3,4]*

[1]Accelerator & Pulsed Power Division, Bhabha Atomic Research Centre, Mumbai 400085, India

[2]Department of Physics, Indian Institute of Technology, Roorkee 247667, Uttarakhand, India

[3]Homi Bhabha National Institute, Mumbai 400094, India

[4]High Pressure &Synchrotron Radiation Physics Division, Bhabha Atomic Research Centre, Mumbai 400085, India

*corresponding author

E-mail: brahma@barc.gov.in

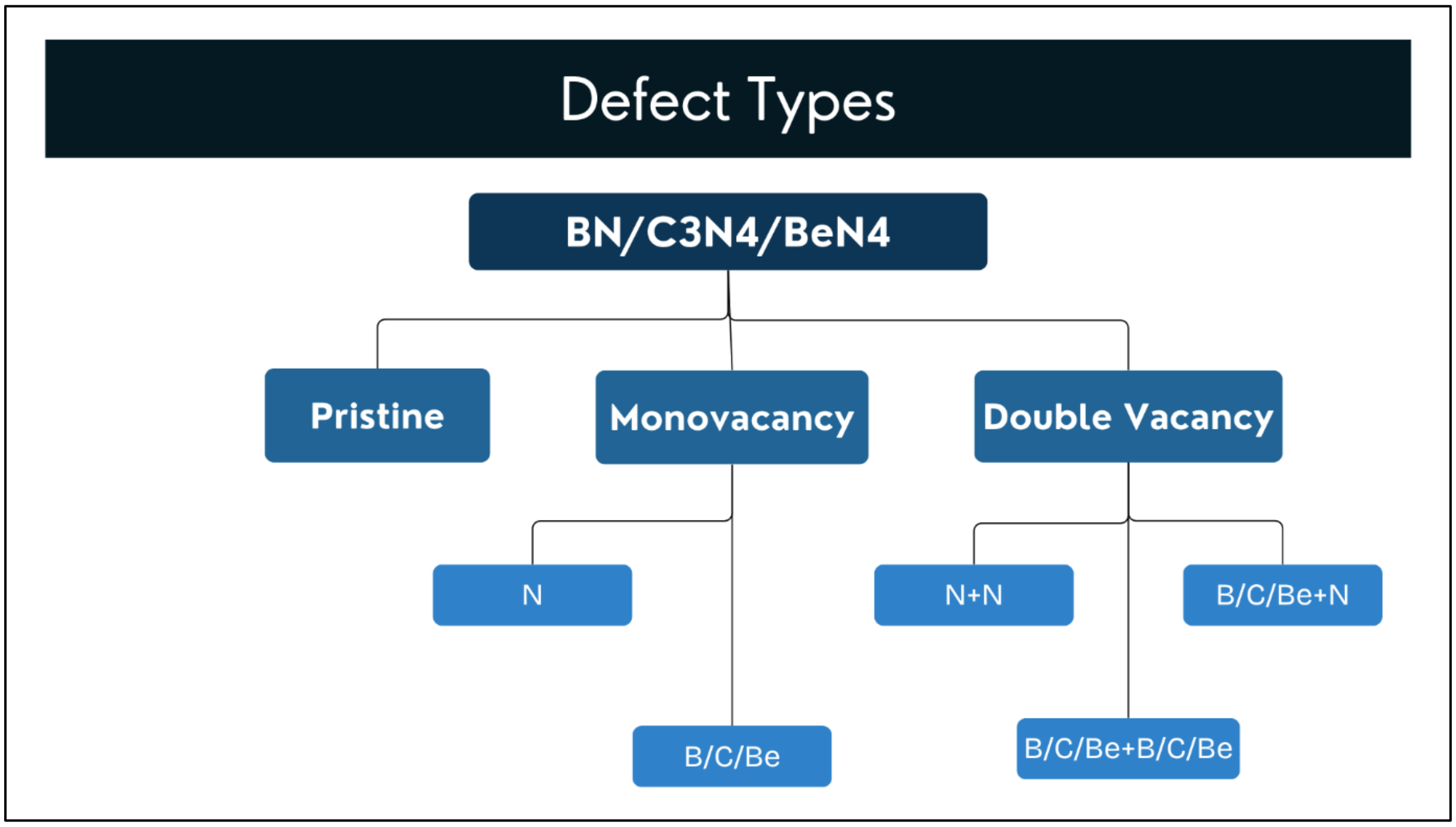

**Figure S1**: Flowchart of all the defect types examined for h-BN/g-C3N4/BeN4

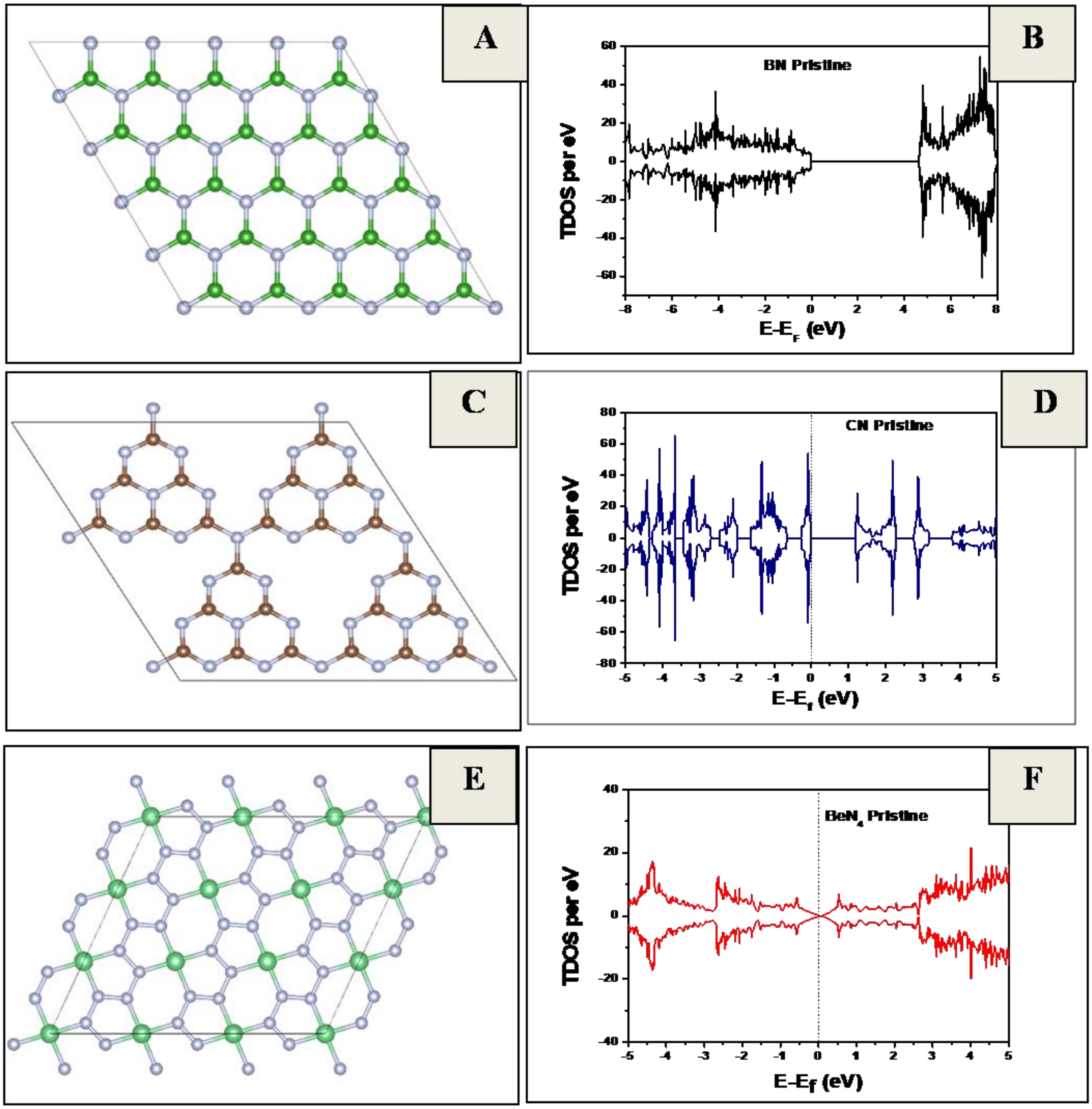

**Figure S2:** Pristine Optimised Structures for (A) Boron Nitride (BN) with B atoms in green and N atoms in grey, (C) Graphitic Carbon Nitride (g-C3N4) with C atoms in Brown and N atoms in grey, (E) Beryllonitrene (BeN4) with Be atoms in light green and N atoms in grey. TDOS plots for (B) Pristine BN, (D) Pristine g-$C_3N_4$ and (F) Pristine $BeN_4$.

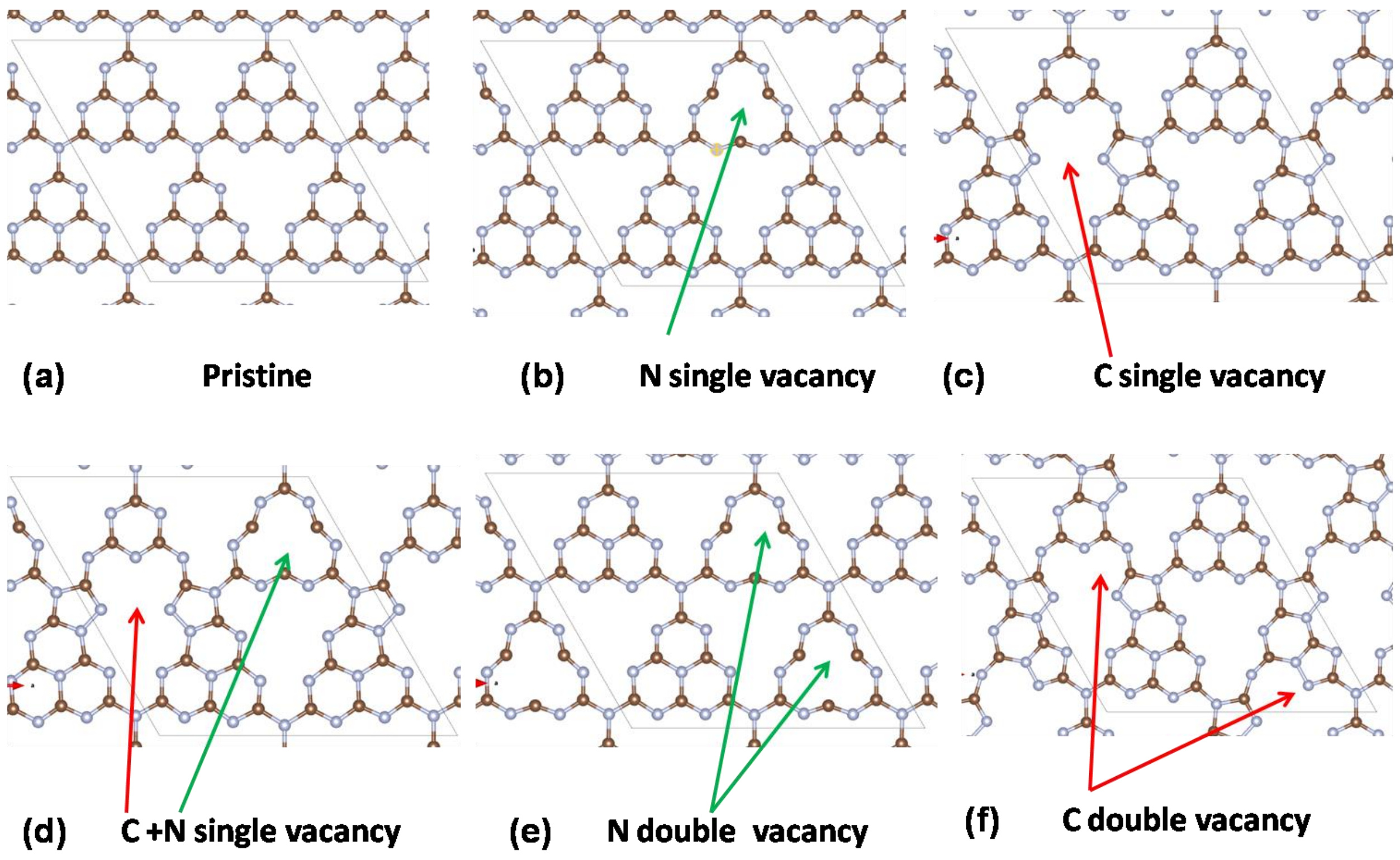

**Figure S3**: Super cells of the optimised structures for Graphitic Carbon Nitride.Carbon atoms are depicted by brown colour and Nitrogen atoms are depicted by transparent colour (a) CNP: relaxed structure of Pristine g-$C_3N_4$; (b) CND1: relaxed structure of g-$C_3N_4$ with a single N atom vacancy defect; (c) CND2: relaxed structure of g-$C_3N_4$ with a single C atom vacancy defect; (d) CND3: relaxed structure of g-$C_3N_4$ with double C + N atoms vacancy defect; (e) CND4: relaxed structure of g-$C_3N_4$ with double N atoms vacancy defect; (f)CND5: relaxed structure of g-$C_3N_4$ with double C atom vacancy defect.

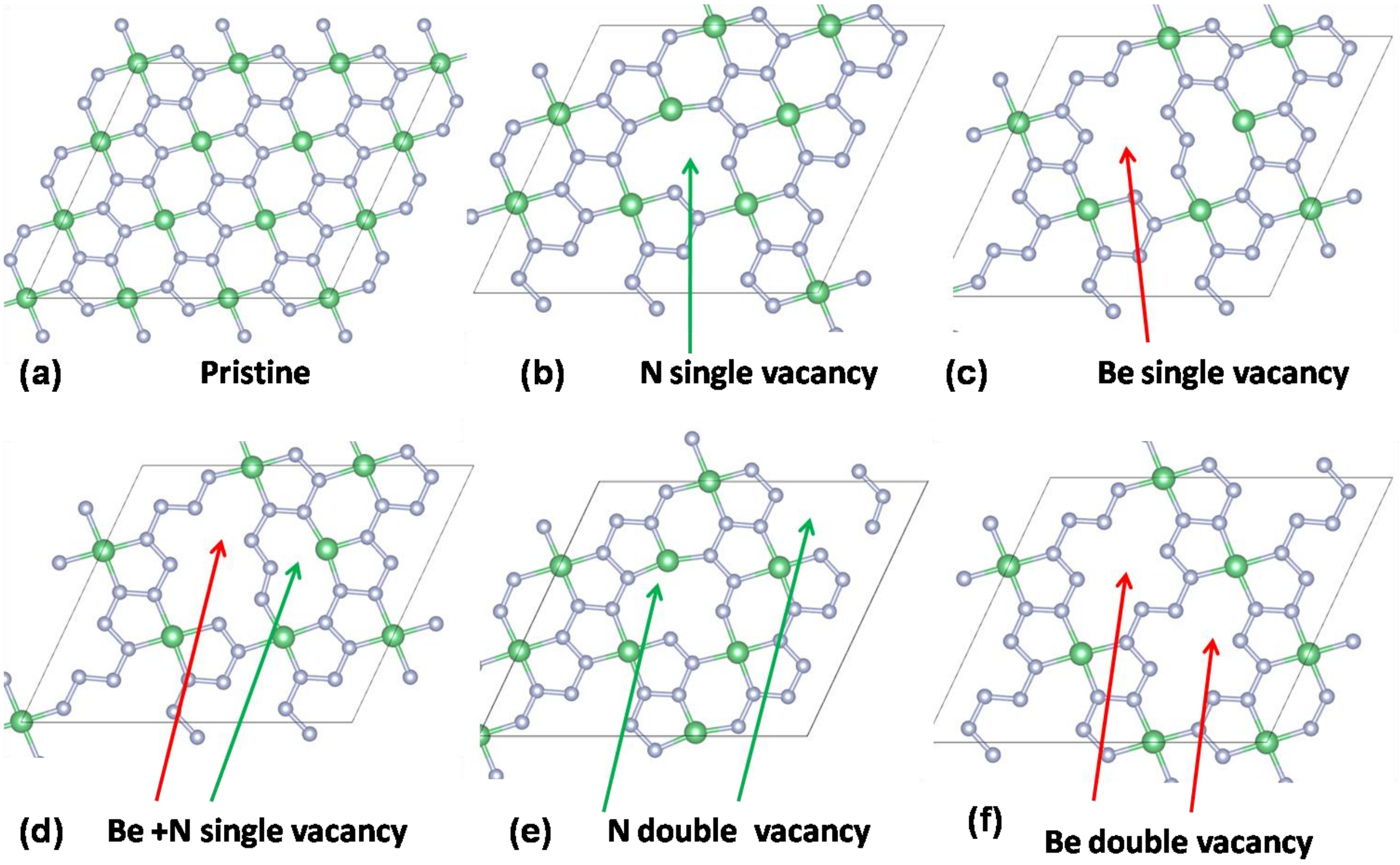

**Figure S4 :** Super cells of the optimised structures for Beryllonitrene. Beryllium atoms are depicted by green colour and Nitrogen atoms are depicted by transparent colour(a) $BeN_4P$: relaxed structure of Pristine $BeN_4$; (b) $BeN_4$ D1: relaxed structure of $BeN_4$ with a single N atom vacancy defect; (c) $BeN_4$ D2: relaxed structure of $BeN_4$ with a single Be atom vacancy defect; (d) $BeN_4$ D3: relaxed structure of $BeN_4$ with double Be + N atoms vacancy defect; (e) $BeN_4$ D4: relaxed structure of $BeN_4$ with double N atoms vacancy defect; (f) $BeN_4$ D5: relaxed structure of $BeN_4$ with double Be atom vacancy defect.

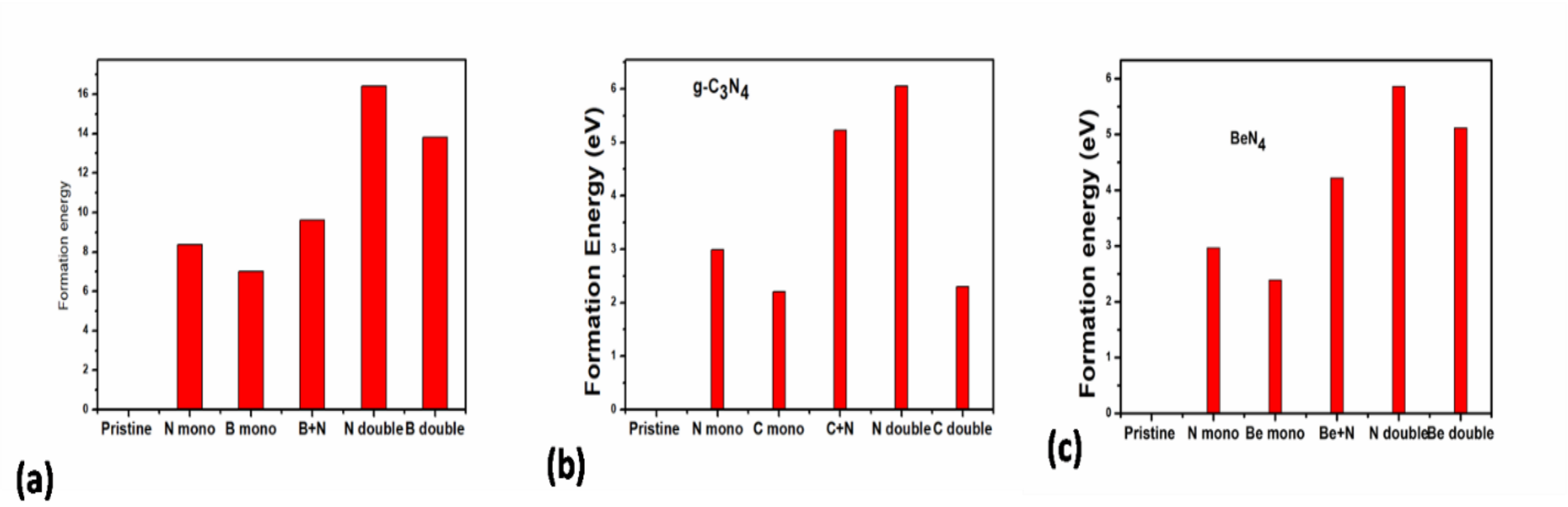

**Figure S5:** Comparison of formation energy of pristine and defective structures of (a) h-BN , (b) g- $C_3N_4$ and (c) $BeN_4$

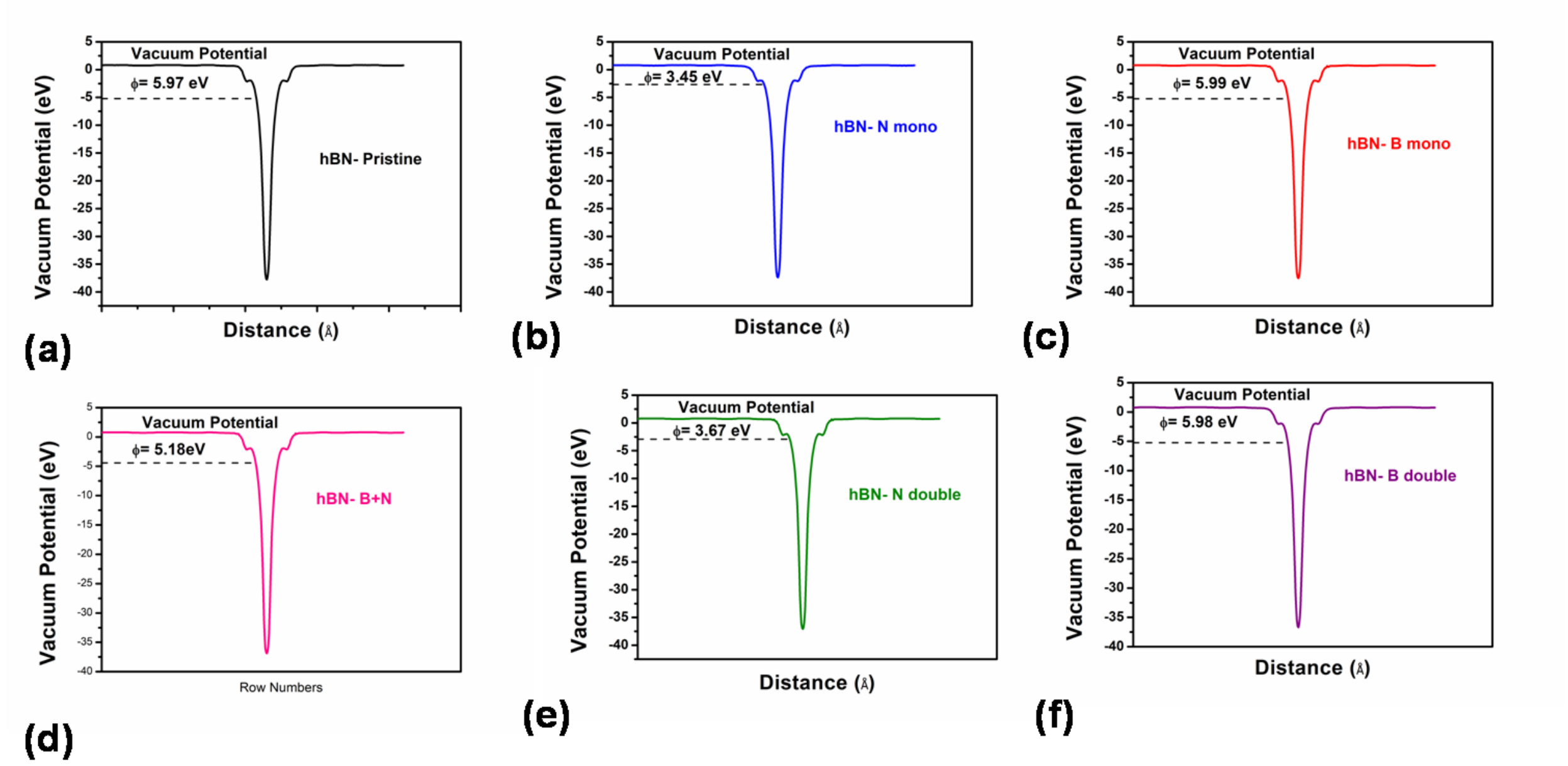

**Figure S6:** Vacuum potential plot forhexagonal Boron Nitride (BN) (a) h-BN Pristine (b) hBN-N mono vacancy defect (c) hBN- B mono vacancy defect, (d) hBN- B+N mono vacancy defect (e)hBN- N double vacancy defect and (f) hBN- B double vacancy defect

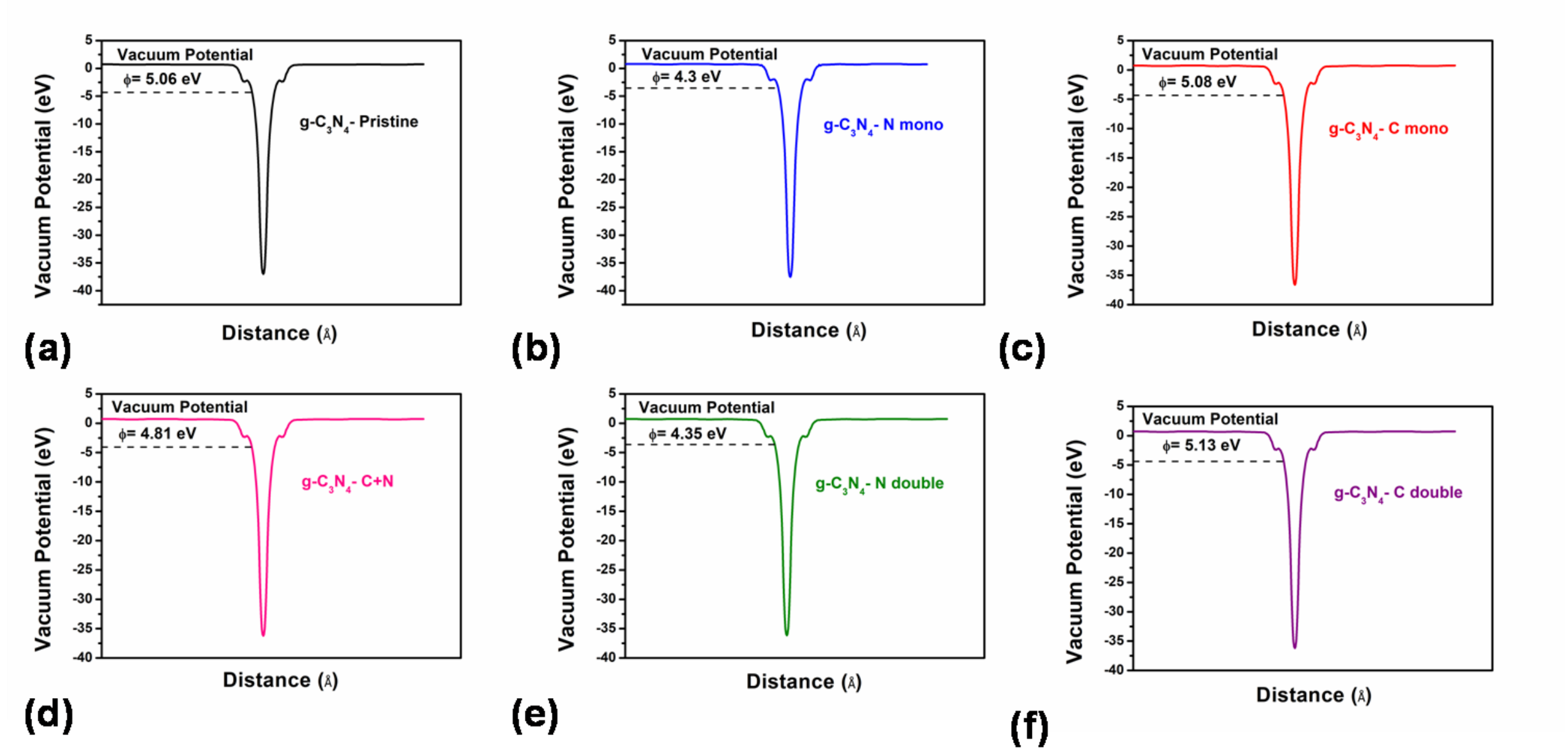

**Figure S7:** Vacuum potential plot for graphitic carbon nitride (g-$C_3N_4$) (a) g-$C_3N_4$ pristine (b) g-$C_3N_4$ -N mono vacancy defect (c) g-$C_3N_4$–C mono vacancy defect, (d) g-$C_3N_4$-C+N mono vacancy defect (e)g-$C_3N_4$- N double vacancy defect and (f) g-$C_3N_4$- C double vacancy defect

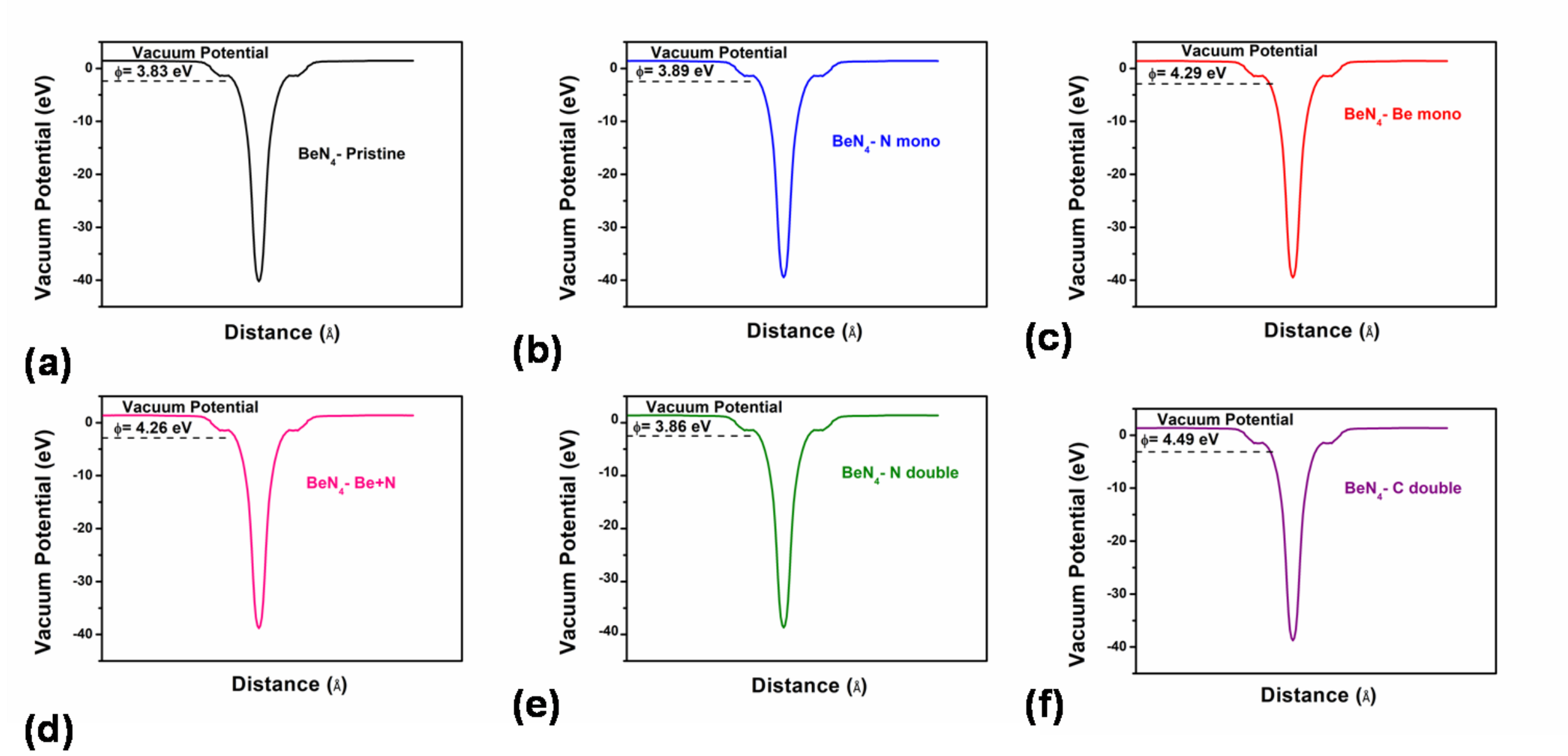

**Figure S8 :** Vacuum potential plot forBeryllonitrene ($BeN_4$) (a) $BeN_4$Pristine (b) $BeN_4$-N mono vacancy defect (c) $BeN_4$– Be mono vacancy defect, (d) $BeN_4$ -Be+N mono vacancy defect (e) $BeN_4$- N double vacancy defect and (f) $BeN_4$- Be double vacancy defect

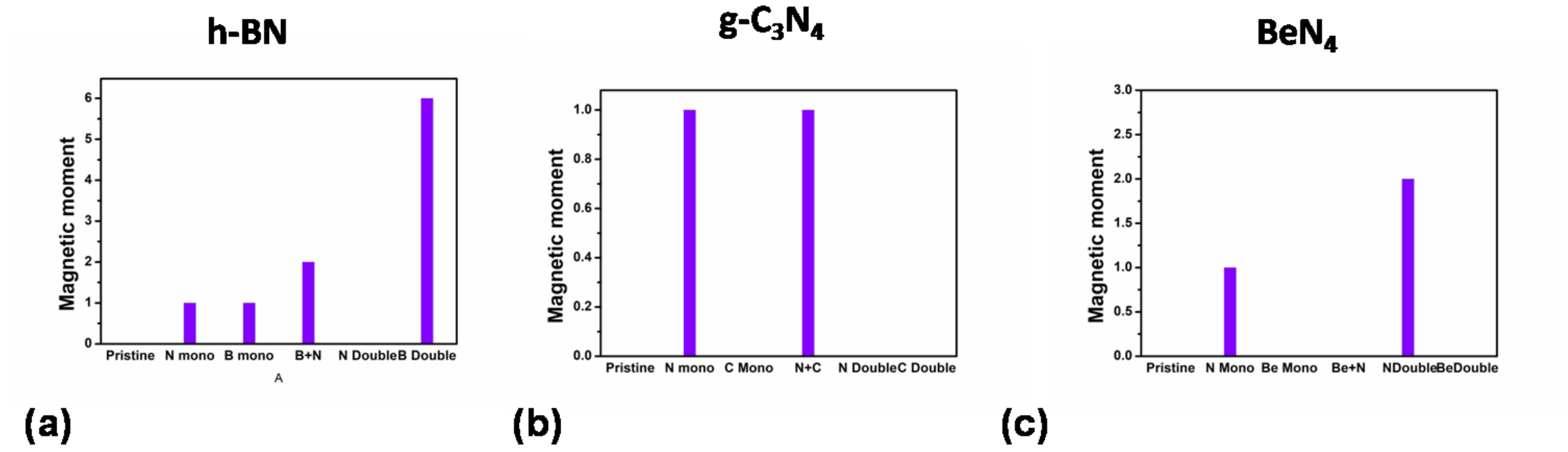

**Figure S9:** Modulation of Magnetic moment values. From left to right; $M_xN_y$ Pristine; $M_xN_y$with a N vacancy; $M_xN_y$ with a M atom vacancy; $M_xN_y$ with a N and M atom double vacancy; $M_xN_y$with a N double vacancy; $M_xN_y$with a M double vacancy where M stands for (a) Boron in h-BN(b) Carbon in g-$C_3N_4$ and (c) Beryllium in $BeN_4$respectively.

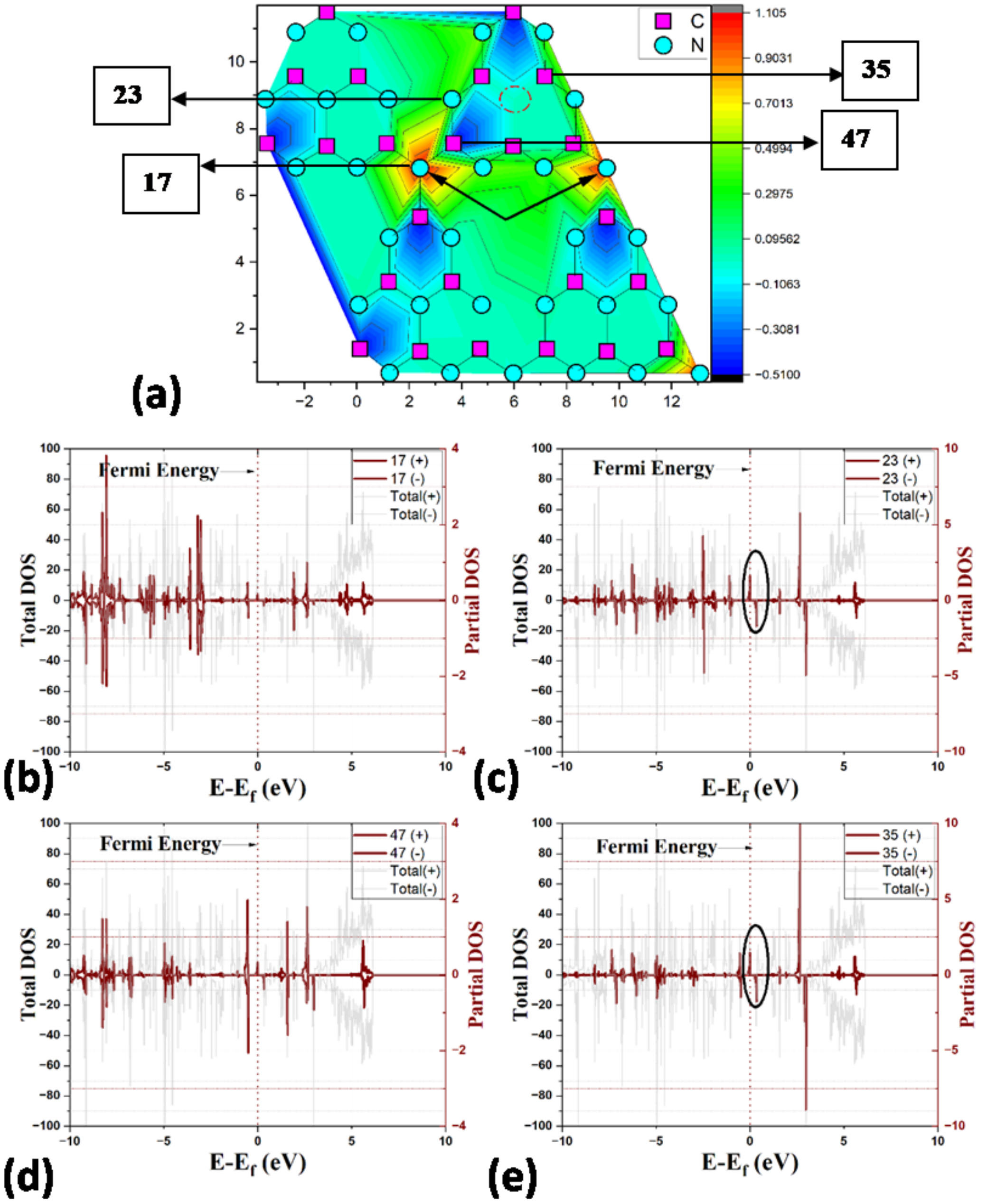

**Figure S10**.:Analysis Plots for Graphitic Carbon Nitride with a nitrogen monovacancy defect. (a) Bader Charge Difference plot for N monovacancy defect. Magenta squares represent Carbon atoms and cyan circles represent Nitrogen atoms in the Super cell. Defect site is marked with a red circle. The plot shows net charge gain at specific nitrogen atoms (bridgehead nitrogen sites) in the super cell (light green and red regions), marked by two black pointers. Atom number 17,23,47,35 marked using black arrows. (b) PDOS plot for Atom 17(Nitrogen atom showing the maximum charge gain) with x axis adjusted according to the Fermi energy. Scale for PDOS

(right) and TDOS (left) are different. (c) PDOS plot for Atom 23 (nearest nitrogen atom to the defect site exhibiting minimal charge gain) with x axis adjusted according to the Fermi energy. Scale for PDOS (right) and TDOS (left) are different. (d) PDOS plot for Atom 47 (Carbon atom near to the defect site exhibiting charge loss) with x axis adjusted according to the Fermi energy. Scale for PDOS (right) and TDOS (left) are different. (e)PDOS plot for Atom 35 (Carbon atom near to the defect site exhibiting charge gain) with x axis adjusted according to the Fermi energy. Scale for PDOS (right) and TDOS (left) are different. Black ellipses highlight the existence of states at the Fermi energy level in partial DOS contribution for the respective atom.

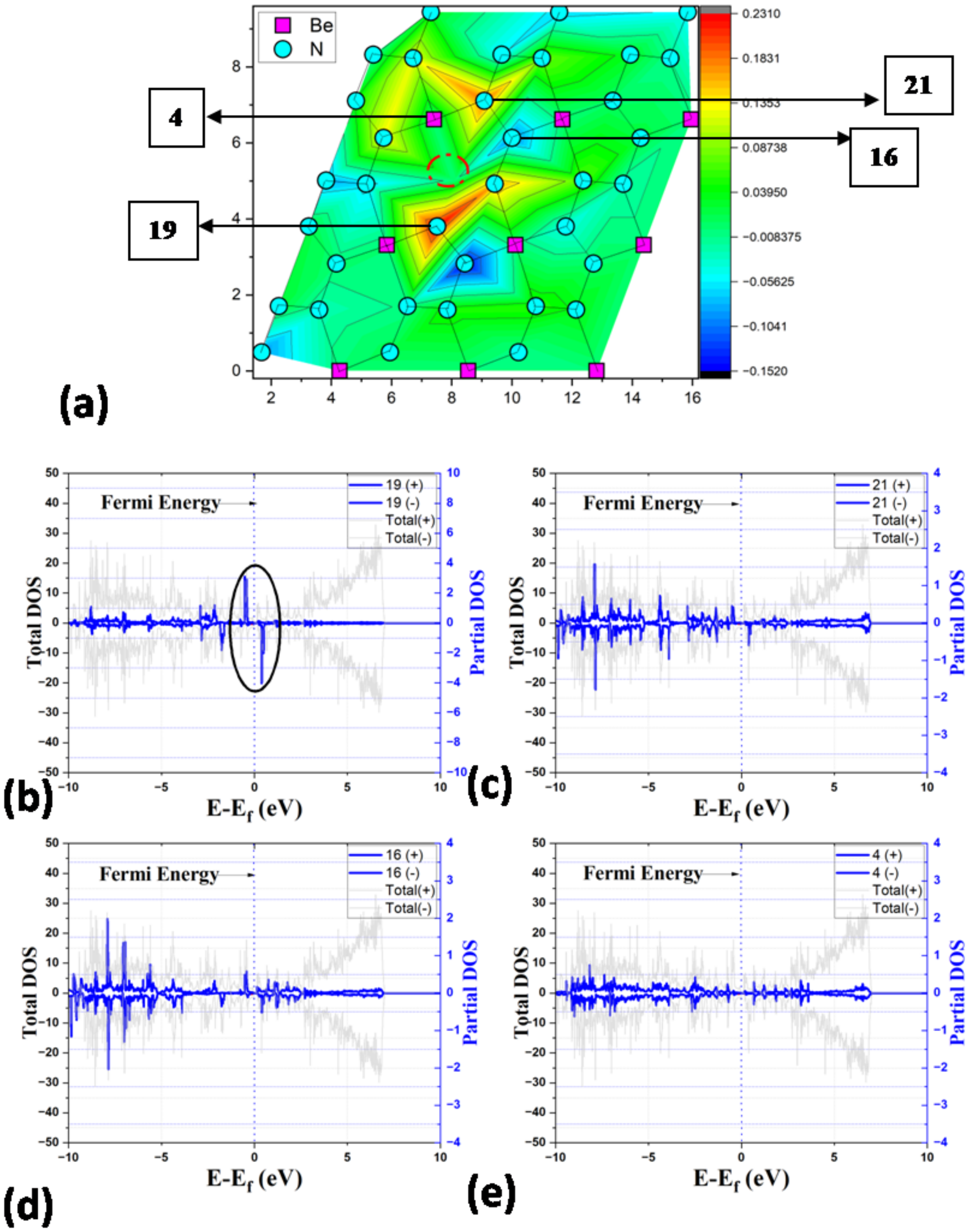

**Figure S11**. Analysis Plots for $BeN_4$ (Beryllonitrene) with a nitrogen monovacancy defect. (a) Bader Charge Difference plot for N monovacancy defect. Magenta squares represent Beryllium atoms and cyan circles represent Nitrogen atoms in the super cell. Defect site is marked with a red circle. The plot shows net charge gain at specific nitrogen atoms in the super cell (light green and red regions), nearby the defect site. Atom numbers4,16,19,21 marked using black arrows. (b) PDOS plot for Atom 19 (Nitrogen atom near the defect site exhibiting charge gain) with x axis adjusted according to the Fermi energy. Scale for PDOS (right) and TDOS (left) are different.(c)PDOS plot for Atom 21 (Nitrogen atom near to the defect site exhibiting minimal charge gain) with x axis adjusted according to the Fermi energy. Scale for PDOS (right) and TDOS (left) are different.(d) PDOS plot for Atom 16 (nearest nitrogen atom to the defect site exhibiting minimal charge loss) with x axis adjusted according to the Fermi energy. Scale for PDOS (right) and TDOS (left) are different.(e) PDOS plot for Atom 4(Beryllium atom nearest to the defect site) with x axis adjusted according to the Fermi energy. Scale for PDOS (right) and TDOS (left) are different. Black ellipses highlight the existence of states at the Fermi energy level in the partial DOS contribution for the respective atom.

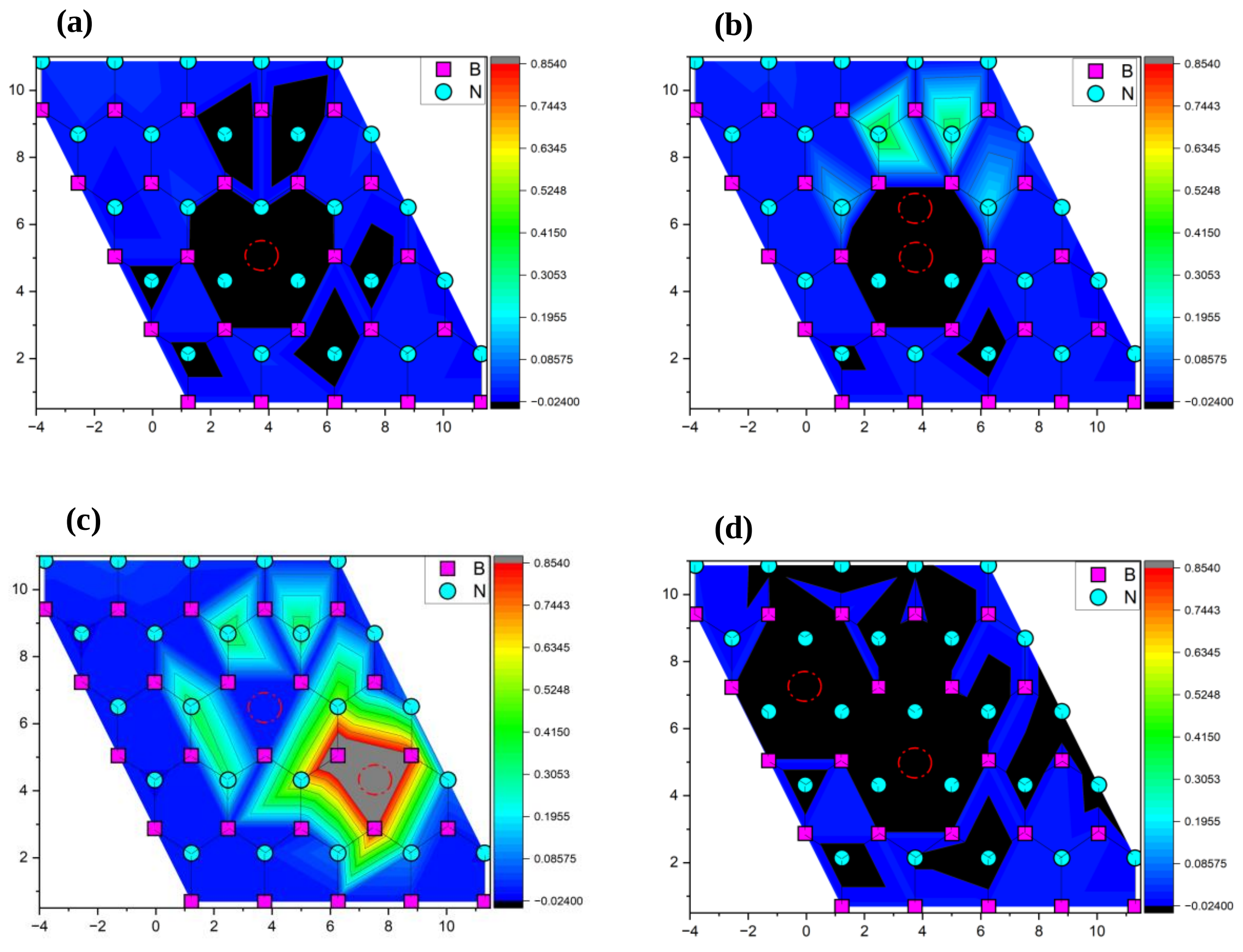

**Figure S12**: Bader charge transfer plots for all defect states of h-BN.Cyan circles denote nitrogen atoms. Pink squares denote boron atoms. The red circles is /are the defect site/sites. (a) Bader charge transfer plot for B mono defect in BN.Net charge loss (black regions) from the entire surface (majoritarily nitrogen atoms) due to loss of electrons of the Boron atom which leads to the slight increase in work function. (b) Bader Charge transfer plot for B+N double defect in BN. Both charge loss (black regions) and charge gain (greenish regions) happening independently due to the B and N vacancy respectively. (c) Bader Charge transfer plot for N double defect in BN. The plot shows net charge gain (light green, red and grey regions) at the neighbouring B and Nitrogen atoms which contribute to the reduction of the work function. (d) Bader Charge transfer plot for B double defect in BN. Net charge loss (black regions) from the entire surface due to loss of electrons of the Boron atoms which leads to the increase in work function.

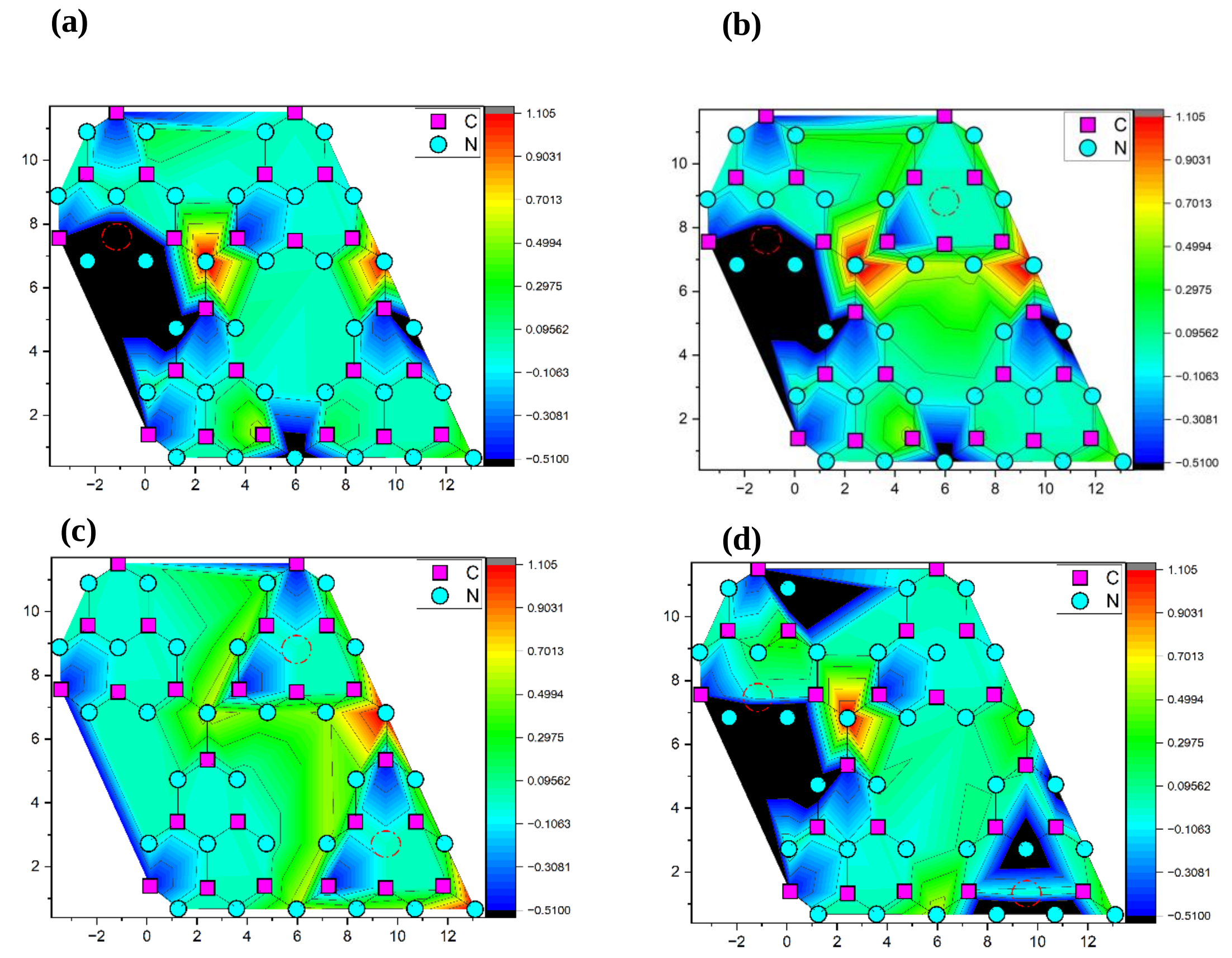

**Figure S13:** Bader charge transfer plots for all defect states of g-$C_3N_4$.Cyan circles denote nitrogen atoms. Pink squares denote carbon atoms. The red circles are the defect sites. (a) Bader charge transfer plot for C mono defect in g-$C_3N_4$.Net charge loss (black regions) from the entire surface (except bridgehead nitrogensites) due to loss of electrons of the carbon atom which leads to the slight increase in work function. (b) Bader charge transfer plot for C+N double defect in g-$C_3N_4$. Both charge loss (black regions) and charge gain (at the bridgeheadnitrogensites) happening independently due to the C and N vacancy respectively. (c) Bader charge transfer plot

for N double defect in g-$C_3N_4$. The plot shows net charge gain (light green and red regions) distributed unevenly on the entire surface which contribute to the reduction of the work function. (d) Bader charge transfer plot for C double defect in g-$C_3N_4$. Net charge loss (black regions) from the entire surface due to loss of electrons of the carbon atoms which leads to the increase in work function.

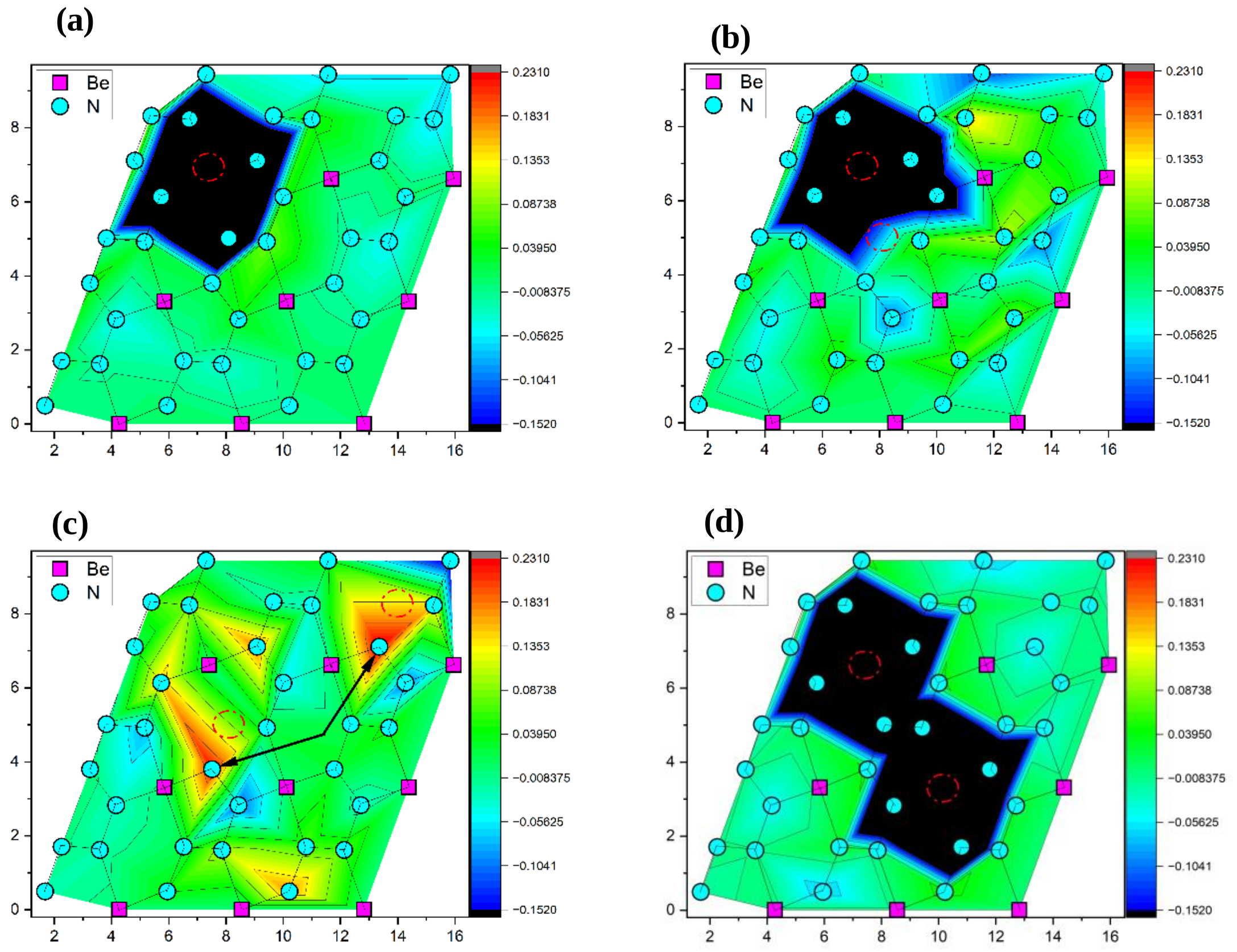

**Figure S14:** Bader charge transfer plots for all defect states of $BeN_4$.Cyan circles denote nitrogen atoms. Pink squares denote beryllium atoms. The red circles are the defect sites. (a) Bader charge transfer plot for Be mono defect in $BeN_4$.Net charge loss (black regions) from the neighbouring nitrogen atoms due to loss of electrons of the beryllium atom which leads to the increase in work function. (b) Bader charge transfer plot for Be +N double defect in $BeN_4$. Net charge loss (back regions) in the neighbouring regions of the defect sites. Trace amounts of charge gains for some nitrogen atoms. (c) Bader charge transfer plot for N double defect in $BeN_4$. The plot shows net charge gain (light green and red regions) at the nearest nitrogen atoms to the defect sites marked by black pointers which contributes to the lower work function (d) Bader charge transfer plot for Be double defect in $BeN_4$. Net charge loss (black regions) especially near the defect sites due to loss of electrons of the beryllium atoms which leads to the increase in work function.

**Table S1:** The simulated optimised bond length of relaxed structures of pristine and vacancy defective structures of h-BN, g-$C_3N_4$ and $BeN_4$

| Material | Defect Type | Bond Type | Bond Length (Å) | Bond Type | Bond Length (Å) |
|---|---|---|---|---|---|
| ***h-BN*** | Pristine | B-N | 1.454 | | - |
| | B mono vacancy | | 1.39 | | - |
| | N mono vacancy | | 1.48 | | - |
| | B+N | | 1.39 | | - |
| | B Double vacancy | | 1.41 | | - |
| | N Double Vacancy | | 1.55 | | - |
| ***g-*** ***$C_3N_4$*** | Pristine | C-N (hexagon ring) | 1.33 | C-N | 1.469 |
| | N mono vacancy | | 1.254 | | 1.44 |
| | C mono vacancy | | 1.37 | | 1.436 |
| | C+N | | 1.25 | | 1.36 |
| | N Double vacancy | | 1.26 | | 1.433 |
| | C Double Vacancy | | 1.386 | | 1.399 |
| ***$BeN_4$*** | Pristine | Be-N | 1.745 | N-N | 1.338 |
| | Be mono vacancy | | 1.757 | | 1.306 |
| | N mono vacancy | | 1.664 | | 1.27 |
| | Be+N | | 1.861 | | 1.28 |
| | Be Double vacancy | | 1.769 | | 1.307 |
| | N Double Vacancy | | 1.691 | | 1.244 |

**Table S2**: Simulated values of band gap and magnetic moment of Pristine and defective structures of h-BN, g-$C_3N_4$ and $BeN_4$

| Material | Defect Type | Band gap (eV) (TDOS analysis) | Band gap (eV) (band structure analysis) | Type of band gap | μ |
|---|---|---|---|---|---|
| ***h-BN*** | Pristine | 4.604 | 4.608 | Indirect | 0 |
| | N mono vacancy | 0.644 | 0.645 | Indirect | 1 |
| | B mono vacancy | 0.158 | 0.185 | Indirect | 1 |
| | B+N | 0.268 | 0.283 | Direct | 2 |
| | N Double vacancy | 0.868 | 0.878 | Direct | 0 |
| | B Double Vacancy | metallic | 0.038 | Indirect | 6 |
| ***g- $C_3N_4$*** | Pristine | 1.189 | 1.189 | Indirect | 0 |
| | N mono vacancy | 0.258 | 0.258 | Indirect | 1 |
| | C mono vacancy | 0.789 | 0.792 | Direct | 0 |
| | C+N | 0.403 | 0.408 | Indirect | 1 |
| | N Double vacancy | 0.06 | 0.259 | Indirect | 0 |
| | C Double Vacancy | 0.809 | 0.808 | Direct | 0 |
| ***$BeN_4$*** | Pristine | 0.114 | 0.114 | Direct | 0 |
| | N mono vacancy | 0.109 | 0.161 | Indirect | 1 |
| | Be mono vacancy | 0.024 | 0.024 | Indirect | 0 |
| | Be+N | 0 | 0 | Metallic | 0 |
| | N Double vacancy | 0.014 | 0 | Metallic | 2 |
| | Be Double Vacancy | 0.049 | 0.148 | indirect | 0 |

**Table S3:** Comparison of work function for studied defective materials with other reported 2D materials.

| Material | Reported Work Function (eV) | Reference |
|---|---|---|
| $PbI_2$ | 4.6 - 5.91 | [S1] |
| $MoS_2$ | 5.23 | [S2] |
| $MoSe_2$ | 4.81 | [S2] |
| $WS_2$ | 5.84 | [S2] |
| $WSe_2$ | 5.37 | [S2] |
| Cs-Graphene | 3.13 | [S3] |
| Graphyne Pristine | 5.39 | [S4] |
| Biphenylene | 4.39 | [S4] |
| Holey Graphyne | 5.56 | [S4] |
| BNP | 5.97 | This Work |
| BND1 | 5.99 | This Work |
| BND2 | 3.45 | This Work |
| BND3 | 5.18 | This Work |
| BND4 | 5.98 | This Work |
| BND5 | 3.67 | This Work |
| CNP | 5.06 | This Work |
| CND1 | 4.3 | This Work |
| CND2 | 5.08 | This Work |
| CND3 | 4.81 | This Work |
| CND4 | 4.35 | This Work |
| CND5 | 5.13 | This Work |
| $BeN_4P$ | 3.83 | This Work |
| $BeN_4D1$ | 3.89 | This Work |
| $BeN_4D2$ | 4.29 | This Work |
| $BeN_4D3$ | 4.26 | This Work |
| $BeN_4D4$ | 3.86 | This Work |
| $BeN_4D5$ | 4.49 | This Work |

**Table S4** :Bader charge distribution of pristine and different defect types of h-BN

| | Charge | | | | | | |
|---|---|---|---|---|---|---|---|
| **Atom #** | **Atom Type** | **Pristine** | **B vacancy** | **N vacancy** | **B+N** | **N double** | **B double** |
| 1 | B | 0 | 0 | 0 | 0 | 0 | 0 |
| 2 | B | 0 | 0 | 0 | 0 | 0 | 0 |
| 3 | B | 0 | 0 | 0 | 0 | 0 | 0 |
| 4 | B | 0 | 0 | 0 | 0 | 0 | 0 |
| 5 | B | 0 | | 0 | | 0 | |
| 6 | B | 0 | 0 | 0 | 0 | 0 | 0 |
| 7 | B | 0 | 0 | 0 | 0 | 0 | 0 |
| 8 | B | 0 | 0 | 0 | 0 | 0.8694 | 0 |
| 9 | B | 0 | 0 | 0 | 0 | 0 | 0 |
| 10 | B | 0 | 0 | 0 | 0 | 0 | 0 |
| 11 | B | 0 | 0 | 0 | 0 | 0 | 0 |
| 12 | B | 0 | 0 | 0 | 0 | 0 | 0 |
| 13 | B | 0 | 0 | 0 | 0 | 0 | 0 |
| 14 | B | 0 | 0 | 0 | 0 | 1.1101 | 0 |
| 15 | B | 0 | 0 | 0 | 0 | 0 | 0 |
| 16 | B | 0 | 0 | 0 | 0 | 0 | 0 |
| 17 | B | 0 | 0 | 0 | 0 | 0 | |
| 18 | B | 0 | 0 | 0 | 0 | 0.8721 | 0 |
| 19 | B | 0 | 0 | 0 | 0 | 0 | 0 |
| 20 | B | 0 | 0 | 0 | 0 | 0 | 0 |
| 21 | B | 0 | 0 | 0 | 0 | 0 | 0 |
| 22 | B | 0 | 0 | 0 | 0 | 0 | 0 |
| 23 | B | 0 | 0 | 0.4553 | 0 | 0 | 0 |
| 24 | B | 0 | 0 | 0 | 0 | 0 | 0 |
| 25 | B | 0 | 0 | 0 | 0 | 0 | 0 |
| 26 | N | 8.0054 | 7.1253 | 8.1365 | 7.4305 | 8.3362 | 7.0883 |
| 27 | N | 8.0055 | 8.0093 | 7.999 | 8.0141 | 8.0011 | 7.1665 |
| 28 | N | 7.9971 | 7.9386 | 8.0101 | 7.9918 | | 7.9451 |
| 29 | N | 8.0029 | 7.9167 | 8.1638 | 8.3907 | 8.3942 | 7.9322 |
| 30 | N | 8.0021 | 7.9564 | 8.0267 | 7.9661 | 8.0084 | 7.942 |
| 31 | N | 8.0076 | 7.9919 | 7.9984 | 8.0058 | 8.015 | 7.1639 |
| 32 | N | 7.9981 | 8.0117 | 8.0132 | 8.008 | 8.0188 | 7.9934 |
| **33** | N | 8.0114 | 7.9341 | 8.0093 | 7.9649 | 8.0867 | 7.943 |
| 34 | N | 8.0042 | **7.0876** | | | | 7.0807 |
| 35 | N | 8.0063 | 8.0036 | 7.9928 | 8.0049 | 8.0222 | 7.9499 |
| 36 | N | 8.0136 | 8.0021 | 7.9904 | 8.0005 | 8.0066 | 7.9952 |
| 37 | N | 8.0041 | 8.0023 | 8.0104 | 8.0139 | 8.0885 | 7.9376 |
| 38 | N | 8.0006 | 7.9906 | 7.9902 | 8.0076 | 8.0135 | 8 |
| 39 | N | 8.0018 | 7.9402 | 8.0227 | 7.9999 | 7.9921 | 7.9384 |
| 40 | N | 7.9965 | 8.0059 | 8.0026 | 7.9988 | 7.9848 | 7.9775 |
| 41 | N | 7.9971 | 7.1403 | 8.8204 | 7.407 | 8.6392 | 7.2131 |
| 42 | N | 8.0058 | 7.9023 | 8.2419 | 8.4402 | 8.407 | 7.9468 |
| 43 | N | 8.0119 | 7.991 | 8.0034 | 7.9861 | 8.0964 | 7.968 |
| 44 | N | 7.9894 | 8.0081 | 8.0032 | 8.0176 | 8.0099 | 7.9887 |
| 45 | N | 8.0008 | 8.0173 | 7.997 | 8.0056 | 8.0083 | 7.9694 |
| 46 | N | 7.9943 | 8.0169 | 8.2477 | 8.1772 | 8.362 | 7.031 |
| 47 | N | 7.9691 | 8.0035 | 8.0074 | 7.9895 | 8.0091 | 7.9377 |
| 48 | N | 7.9883 | 8.0018 | 7.9941 | 8.0028 | 8.084 | 7.9968 |
| 49 | N | 8.0027 | 8.0111 | 8.8548 | 8.1995 | 8.5662 | 7.9432 |
| 50 | N | 7.9835 | 7.9914 | 8.0089 | 7.9769 | 7.9982 | 7.9517 |
| | | | | | | | |

**Table S5** : Bader charge distribution of pristine and different defect types of Graphitic Carbon nitride

| | | **Charge** | | | | | |
|---|---|---|---|---|---|---|---|
| Atom # | **Atom type** | **Pristine** | **N vacancy** | **C vacancy** | **C+N** | **N double** | **C double** |
| 1 | N | 7.8993 | 7.88 | 7.8719 | 7.866 | 7.8783 | 7.8368 |
| 2 | N | 7.8993 | 7.8925 | 7.9909 | 7.9655 | 7.8783 | 8.1456 |
| 3 | N | 7.8993 | 7.8924 | 7.9018 | 7.8817 | | 7.0026 |
| 4 | N | 7.8993 | | 7.847 | | | 7.8716 |
| 5 | N | 8.1141 | 8.1442 | 7.2085 | 7.2519 | 8.1518 | 7.2553 |
| 6 | N | 8.1141 | 8.1132 | 8.1535 | 8.1419 | 8.1518 | 8.0603 |
| 7 | N | 8.1141 | 8.0175 | 8.1335 | 8.1677 | 8.481 | 8.2025 |
| 8 | N | 8.1141 | 8.6017 | 8.1499 | 8.4704 | 8.481 | 8.079 |
| 9 | N | 8.0251 | 8.027 | 8.0885 | 8.0924 | 8.0452 | 8.0752 |
| 10 | N | 8.0251 | 8.0442 | 6.5733 | 6.5801 | 8.0452 | 6.5872 |
| 11 | N | 8.0251 | 8.0734 | 8.1368 | 8.1329 | 8.5522 | 6.6936 |
| 12 | N | 8.0251 | 8.4341 | 8.0449 | 8.6071 | 8.5522 | 8.0803 |
| 13 | N | 8.1108 | 8.1104 | 8.0561 | 8.0985 | 8.0718 | 8.0429 |
| 14 | N | 8.1108 | 8.0732 | 8.0508 | 8.0389 | 8.0718 | 8.1682 |
| 15 | N | 8.1108 | 8.044 | 8.1443 | 8.1701 | 8.4577 | 8.2345 |
| 16 | N | 8.1108 | 8.5546 | 8.0264 | 8.3878 | 8.4577 | 8.012 |
| 17 | N | 6.6849 | 6.6781 | 5.8688 | 5.9652 | 7.2798 | 7.3287 |
| 18 | N | 6.6849 | 7.7876 | 7.9394 | 7.9391 | 7.2798 | 7.8973 |
| 19 | N | 6.6849 | 7.7786 | 7.1725 | 7.2353 | 7.8823 | 7.1644 |
| 20 | N | 6.6849 | 7.7877 | 7.8859 | 7.9557 | 7.8823 | 7.1263 |
| 21 | N | 8.0343 | 8.0647 | 8.1942 | 8.1889 | 8.0381 | 8.1018 |
| 22 | N | 8.0343 | 8.0175 | 8.0394 | 7.9812 | 8.0381 | 8.1291 |
| 23 | N | 8.0343 | 8.1133 | 8.0516 | 8.0534 | 8.5733 | 8.1702 |
| 24 | N | 8.0343 | 8.4422 | 8.0077 | 8.3918 | 8.5733 | 8.0685 |
| 25 | N | 8.0481 | 8.0415 | 8.142 | 8.152 | 8.0528 | 8.1698 |
| 26 | N | 8.0481 | 8.0438 | 8.2894 | 8.3567 | 8.0528 | 6.4949 |
| 27 | N | 8.0481 | 8.0696 | 7.2192 | 7.1926 | 8.4987 | 7.3088 |
| 28 | N | 8.0481 | 8.5276 | 8.0894 | 8.525 | 8.4987 | 8.0977 |
| 29 | N | 8.065 | 8.0486 | 8.1078 | 8.0912 | 8.0572 | 8.1238 |
| 30 | N | 8.065 | 8.087 | 6.6158 | 6.6551 | 8.0572 | 6.6041 |
| 31 | N | 8.065 | 8.061 | 8.0349 | 8.0531 | 8.4872 | 6.7758 |
| 32 | N | 8.065 | 8.5277 | 8.0762 | 8.695 | 8.4872 | 8.0702 |
| 33 | C | 0 | 0 | 0 | 0 | 0 | 0 |
| 34 | C | 0 | 0 | 0 | 0 | 0 | 0.4101 |
| 35 | C | 0 | 0 | 0 | 0 | 0 | 0 |
| 36 | C | 0 | 0 | 0 | 0 | 0 | 0 |
| 37 | C | 0.5091 | 0 | 0 | 0 | 0.4925 | 0 |
| 38 | C | 0.5091 | 0.5106 | 0 | 0 | 0.4925 | 0 |
| 39 | C | 0.5091 | 0 | 0 | 0 | 0 | 0.6226 |
| 40 | C | 0.5091 | 0 | 0.5201 | 0.4738 | 0 | 0.5528 |
| 41 | C | 0 | 0 | 0.6424 | 0.616 | 0 | 0 |
| 42 | C | 0 | 0 | 0 | 0 | 0 | 0 |
| 43 | C | 0 | 0 | 0 | 0 | 0 | 0 |
| 44 | C | 0 | 0 | 0 | 0 | 0 | 0 |
| 45 | C | 0.5092 | 0 | 0 | 0 | 0 | 0 |
| 46 | C | 0.5092 | 0 | 0 | 0 | 0 | 0 |
| 47 | C | 0.5092 | 0.5106 | 0.7251 | 0.6259 | 0 | 0.4355 |
| 48 | C | 0.5092 | 0 | 0 | 0 | 0 | 0 |
| 49 | C | 0 | 0 | 0 | 0 | 0 | 0 |
| 50 | C | 0 | 0 | | | 0 | |
| 51 | C | 0 | 0 | 0 | 0 | 0 | |
| 52 | C | 0 | 0 | 0 | 0 | 0 | 0 |
| 53 | C | 0 | 0 | 0 | 0 | 0 | 0 |
| 54 | C | 0 | 0 | 0 | 0 | 0 | 0 |
| 55 | C | 0 | 0 | 0 | 0 | 0 | 0 |
| 56 | C | 0 | 0 | 0 | 0 | 0 | 0 |

**Table S6:** Bader charge distribution of pristine and different defect types of $BeN_4$

| | | Charge | | | | | |
|---|---|---|---|---|---|---|---|
| **Atom #** | **Atom Type** | **Pristine** | **N vacancy** | **Be vacancy** | **Be+N** | **N double** | **Be double** |
| 1 | Be | 0 | 0 | 0 | 0 | 0 | 0 |
| 2 | Be | 0 | 0 | 0 | 0 | 0 | 0 |
| 3 | Be | 0 | 0 | 0 | 0 | 0 | |
| 4 | Be | 0 | 0 | 0 | 0 | 0 | 0 |
| 5 | Be | 0 | 0 | | | 0 | 0 |
| 6 | Be | 0 | 0 | 0 | 0 | 0 | 0 |
| 7 | Be | 0 | 0 | 0 | 0 | 0 | |
| 8 | Be | 0 | 0 | 0 | 0 | 0 | 0 |
| 9 | Be | 0 | 0 | 0 | 0 | 0 | 0 |
| 10 | N | 5.4881 | 5.5221 | 5.4687 | 5.5025 | 5.4846 | 5.5385 |
| 11 | N | 5.5116 | 5.512 | 5.4623 | 5.4569 | 5.4839 | 5.4602 |
| 12 | N | 5.4868 | 5.618 | 5.0149 | 5.05 | 5.637 | 5.008 |
| 13 | N | 5.4881 | 5.4822 | 5.4848 | 5.4808 | 5.6593 | 5.5388 |
| 14 | N | 5.5116 | 5.5176 | 5.5128 | 5.5663 | 5.7509 | 5.4602 |
| 15 | N | 5.4868 | 5.4443 | 5.4709 | 5.4593 | 5.5138 | 5.0077 |
| 16 | N | 5.4994 | 5.4014 | 5.5498 | 5.2344 | 5.4457 | 5.5441 |
| 17 | N | 5.5513 | 5.4849 | 5.489 | 5.432 | 5.5895 | 5.4804 |
| 18 | N | 5.4954 | 5.3436 | 5.5317 | 5.3921 | 5.3842 | 4.9751 |
| 19 | N | 5.4994 | 5.7299 | 5.5183 | 5.4754 | 5.7203 | 5.5438 |
| 20 | N | 5.5076 | 5.5211 | 5.5117 | 5.5169 | 5.548 | 5.4325 |
| 21 | N | 5.4954 | 5.6805 | 4.9721 | 4.9624 | 5.6733 | 4.9736 |
| 22 | N | 5.4476 | 5.4533 | 5.4363 | 5.5315 | 5.4193 | 5.4754 |
| 23 | N | 5.4738 | 5.495 | 5.4789 | 5.5327 | 5.3679 | 5.47 |
| 24 | N | 5.5451 | 5.4927 | 5.482 | 5.494 | 5.3744 | 5.529 |
| 25 | N | 5.4476 | 5.4919 | 5.5182 | 5.5048 | 5.4567 | 5.4739 |
| 26 | N | 5.4738 | 5.4923 | 5.5232 | 5.5245 | 5.475 | 5.4674 |
| 27 | N | 5.5451 | 5.4527 | 5.5425 | 5.5528 | 5.525 | 5.5294 |
| 28 | N | 5.4818 | 5.6057 | 5.0207 | 5.0222 | 5.5712 | 4.9572 |
| 29 | N | 5.4671 | 5.445 | 5.4552 | 5.4251 | 5.4482 | 5.4602 |
| 30 | N | 5.4918 | 5.4273 | 5.5172 | 5.5112 | 5.4111 | 5.5042 |
| 31 | N | 5.4818 | 5.5372 | 5.4875 | 5.5485 | 5.5465 | 4.959 |
| 32 | N | 5.4671 | 5.4985 | 5.4626 | 5.5188 | | 5.46 |
| 33 | N | 5.4918 | 5.5604 | 5.4688 | 5.5965 | 5.5116 | 5.5075 |
| 34 | N | 5.4782 | 5.6395 | 5.5602 | 5.5493 | 5.501 | 4.9702 |
| 35 | N | 5.4712 | 5.5497 | 5.5249 | 5.6175 | 5.4391 | 5.4973 |
| 36 | N | 5.5386 | 5.5905 | 5.5053 | 5.556 | 5.6422 | 5.5398 |
| 37 | N | 5.5402 | | 4.9797 | | | 5.0853 |
| 38 | N | 5.4712 | 5.4657 | 5.4507 | 5.5297 | 5.4865 | 5.4988 |
| 39 | N | 5.5386 | 5.4838 | 5.528 | 5.4871 | 5.5185 | 5.5422 |
| 40 | N | 5.5059 | 5.5426 | 5.5122 | 5.4475 | 5.5707 | 5.5331 |
| 41 | N | 5.5409 | 5.5223 | 5.4886 | 5.4354 | 5.5861 | 5.5044 |
| 42 | N | 5.5167 | 5.5074 | 5.4499 | 5.5142 | 5.6602 | 5.5137 |
| 43 | N | 5.506 | 5.4937 | 5.5777 | 5.5129 | 5.566 | 5.5344 |
| 44 | N | 5.5398 | 5.4562 | 5.5158 | 5.5262 | 5.5198 | 5.5094 |
| 45 | N | 5.5167 | 5.5388 | 5.5268 | 5.5324 | 5.5123 | 5.515 |

## References:

[S1]Chen, Hongfei, Hejin Yan, and Yongqing Cai. "Effects of defect on work function and energy alignment of PbI2: implications for solar cell applications." *Chemistry of Materials* 34.3 (2022): 1020-1029.

[S2]Pawar, Ravinder, and Akanksha Ashok Sangolkar. "Density functional theory based HSE06 calculations to probe the effects of defect on electronic properties of monolayer TMDCs." *Computational and Theoretical Chemistry* 1205 (2021): 113445.

[S3] Legesse, M.; El Mellouhi, F.; Bentria, E.T.; Madjet, M.E.; Fisher, T.S.; Kais, S.; Alharbi, F.H. Reduced work function of graphene by metal adatoms. Appl

[S4] Sarkar S. G.; Jethawa U.; Sanyal G.; Chakraborty B. Work Function Modulation in 2D Carbon Allotropes via Defect Engineering for Field Emission Applications: A DFT Analysis *ACS Appl. Electron. Mater*. **2024**, 6, 12, 8898–8911.